\documentclass[aps,prd,superscriptaddress,nofootinbib,eqsecnum,twocolumn]{revtex4-1}

\pdfoutput=1

\usepackage{amsfonts}
\usepackage{amsmath}
\usepackage{amssymb}
\usepackage{graphicx,color}
\usepackage{float}
\usepackage{hyperref}
\usepackage{subfigure}
\usepackage{mathtools,slashed}

\usepackage{longtable}
\usepackage{dcolumn}
\usepackage{bm}
\usepackage{appendix}
\usepackage{multirow}
\usepackage{color}
\usepackage{soul}
\usepackage[normalem]{ulem}

\begin{document}

\title{Warm bounce in loop quantum cosmology and the prediction for the duration of inflation}

\author{L. N. Barboza} 
\email{ln\textunderscore barboza@id.uff.br}
\affiliation{Instituto de F\'{\i}sica, Universidade Federal Fluminense, 
Avenida General Milton Tavares de Souza s/n, Gragoat\'a, 24210-346 Niter\'oi, 
Rio de Janeiro, Brazil}

\author{L. L. Graef} 
\email{leilagraef@id.uff.br}
\affiliation{Instituto de F\'{\i}sica, Universidade Federal Fluminense, 
Avenida General Milton Tavares de Souza s/n, Gragoat\'a, 24210-346 Niter\'oi, 
Rio de Janeiro, Brazil}

\author{Rudnei O. Ramos} 
\email{rudnei@uerj.br}
\affiliation{Departamento de F\'{\i}sica Te\'orica, Universidade do Estado do 
Rio de Janeiro, 20550-013 Rio de Janeiro, RJ, Brazil}

\begin{abstract}

We study and estimate probabilistic predictions for the
duration of the preinflationary and slow-roll phases after the bounce in loop quantum cosmology, determining how the presence of radiation in the pre-bounce phase affects these results. 
We present our analysis for different classes of inflationary
potentials that include the monomial power-law chaotic type of potentials,
namely, for the quadratic, quartic  and sextic potentials and also for a
Higgs-like symmetry-breaking potential, considering different values for
the vacuum expectation value in the latter case. We obtain the
probability density function for the number of inflationary \textit{e}-folds
and for other relevant quantities for each model and produce
probabilistic results drawn from these distributions. This study
allows us to discuss under which conditions  each model  could either
eventually lead to observable signatures in the spectrum of the cosmic
microwave background, or be excluded for not predicting a
sufficient amount of accelerated expansion. The effect of radiation on the 
predictions for each model is explicitly quantified.
The obtained results indicate that the number of inflationary \textit{e}-folds 
in loop quantum cosmology is not \textit{a priori} an arbitrary number, but can in 
principle be a predictable quantity, even though the results are 
dependent on the model and the amount of radiation in the Universe
prior to the start of the inflationary regime. 

\end{abstract}

\maketitle

\section{Introduction}
\label{intro}

Inflation is the current paradigm for the early Universe
cosmology.\footnote{Although inflation is the current  paradigm for the early Universe cosmology, it is worth mentioning that there are alternative ideas~\cite{Lilley:2015ksa,Creminelli:2010ba,Brandenberger:2009jq}, like several bouncing models, which can agree with current cosmological observations as well as inflation does.} The inflationary scenario was developed before a majority of current data was recorded. Inflation is in good agreement with the predictions coming
from the cosmic microwave background (CMB) spectrum  and explains the
origin of inhomogeneities present in the primordial Universe, which
led to the formation of large-scale structures. Thus, although
fine-tunings of the constants are necessary and  appropriate choices
of potentials have to be made, this is a very predictive
scenario. Inflation is a good candidate for solving some of the puzzles in
the standard big bang cosmology, such as the horizon and
flatness - problems~\cite{Guth:1980zm,Linde:2007fr,Vazquez:2018qdg}. Despite its success, 
the idea of inflation alone does not address the important issue of
extending general relativity (GR) beyond its limit of applicability,
which is associated with the big bang singularity problem.  Apart from
this problem, one should consider in the space of classic solutions
for GR those solutions that exhibit sufficient inflation to account
for the current
observations~\cite{Gibbons:1986xk,Kofman:2002cj,Brandenberger:2011gk}. This
motivates an investigation of the probability of a sufficient amount
of inflation in a cosmological model. In this endeavor, one is plagued
with problems, such as the difficulty in defining a measure to
calculate probabilities in GR and finding the starting point for
counting \textit{e}-folds in the presence of a
singularity~\cite{Chen:2015yua,Gibbons:2006pa}. These problems have
received a lot of attention in recent years~\cite{Hawking:1987bi}. In
order to better address these issues, we consider here a
nonperturbative quantum gravity theory independent of the GR
background, that is, loop quantum gravity
(LQG)~\cite{Ashtekar:2011ni,Barrau:2013ula,Agullo:2016tjh,Bojowald:2006da,Ashtekar:2003hd,Ashtekar:2006rx}.

Loop quantum cosmology (LQC) is the reduced
version of LQG~\cite{Ashtekar:2003hd}, which uses the symmetries considered in cosmology. It 
uses the so-called Ashtekar variables and its quantization is
obtained from holonomies of the connections and fluxes of the
densitized triads. However, taking into account such quantum geometric
effects in cosmological models, while Einstein's equations maintain an
excellent degree of approximation at low curvature,  in the Planck
regime they undergo major changes. In LQC the big bang singularity is 
naturally resolved and replaced by a  bounce due to repulsive quantum geometry
effects~\cite{Ashtekar:2011ni,Ashtekar:2009mm}.  In LQC, for
matter that satisfies the normal conditions of energy, whenever a
curvature invariant grows at the Planck scale the effects of quantum
geometry dilute it, thus resolving the singularities of
GR~\cite{Ashtekar:2011ni}.

Within the community of LQC there is a lively debate on the
naturalness of the emergence of an inflationary phase after the
bounce, and following this line, there is a search for  the most
probable number of inflationary \textit{e}-folds predicted by a
model~\cite{Barrau:2020nek}. {}First of all, in addressing this
question the measure problem  is something that requires quite some
attention, given that there is no consensus on how to establish the
initial conditions necessary to obtain the dynamics of the models and 
compute probabilities. Since there is no direct observational information
from the initial conditions of the Universe, one has to consider all
possible initial conditions  to draw conclusions about the probability
of an inflationary phase~\cite{Bedic:2018gqu}.  

Beginning from the GR context, the possibility of using the Liouville
measure as a candidate to calculate the probability was discussed by
Gibbons \textit{et al.}~\cite{Gibbons:1986xk}. However, in the flat {}Friedmann-Lema\^itre-Robertson-Walker model this total
Liouville measure is infinite, requiring a regularization
scheme~\cite{Chen:2015yua,Gibbons:2006pa}. Besides that, there is a
huge discrepancy between the probability  estimated by Gibbons and
Turok~\cite{Gibbons:2006pa} and the results obtained, for example, by
Linde~\cite{Linde:2007fr}.

In LQC, since the singularity of the big bang is  solved and it is
replaced by a (quantum)
bounce~\cite{Ashtekar:2006rx,Ashtekar:2006wn,Ashtekar:2007em},  a
regular surface can be used to introduce the structure needed to
specify a Liouville measure (see also Refs.~\cite{Graef:2018ulg,Bedic:2018gqu} 
for extensions of this approach).  The problem of
making a measurement present in GR~\cite{Gibbons:1986xk} is naturally
resolved in LQC~\cite{Corichi:2010zp}. In the absence of the
singularity, an {\it a priori} probability for a sufficiently long
slow-roll inflation phase can then be obtained. However, also in the
context of LQC, different approaches have been advocated. Ashtekar and Sloan~\cite{Ashtekar:2011rm} argued that a natural
measure can be  implemented in LQC and proposed a Planck surface scale,
with which probabilities can be calculated. The approach advocated in Ref.~\cite{Ashtekar:2011rm} does not agree with the one suggested in Refs.~\cite{Linsefors:2013cd,Linsefors:2014tna,Martineau:2017sti,Bolliet:2017czc}. Despite the current debate, many works have consistently shown that in
LQC models with a kinetic-energy-dominated bounce an inflationary
phase almost inevitably sets in (see, e.g., 
Refs.~\cite{Zhu:2017jew,Shahalam:2017wba,Li:2018fco,Sharma:2018vnv,Li:2019ipm,Shahalam:2019mpw}). 

In addition to showing the naturalness of inflation, it is  important to
investigate the most probable number of inflationary \textit{e}-folds predicted
by these models. As it is well known~\cite{Guth:1980zm}, the
inflationary phase must last at least around $60$ or so \textit{e}-folds in
order to solve the main problems that inflation is expected to. On the
other hand, another important question is whether the quantum bounce
and subsequent preinflationary phase can leave  observational
signatures that can be observed in current and forthcoming
experiments~\cite{Agullo:2013ai,Barrau:2016nwy}. As shown in
Ref.~\cite{Agullo:2013ai}, the bounce and preinflationary dynamics
leaves  imprints on the spectrum of the CMB. In
Ref.~\cite{Zhu:2017jew} it was shown that in LQC models,  in order to
be consistent with observations, the Universe must have expanded at
least around $141$ \textit{e}-folds from the bounce until now. This is so
because LQC can lead to scale-dependent features in the CMB, and the fact that we do not observe them today means that they must have been well
diluted by the post-bounce expansion of the Universe. By comparing that
total number of expansion of the Universe to the minimum number of
inflationary \textit{e}-folds required (added to the typical 60 \textit{e}-folds 
from the end of inflation until today), this implies an extra number of
inflationary \textit{e}-folds in LQC, given by $\delta N \sim
21$~\cite{Zhu:2017jew}. On the other hand, if the number of extra
inflationary \textit{e}-folds is much higher than this value the features 
imprinted in the CMB spectrum due to the LQC effects are too  diluted, 
and in this case LQC cannot be directly tested even by forthcoming experiments. This
motivates a deep investigation of the most probable number of \textit{e}-folds
in models of LQC.  The most probable number of inflationary \textit{e}-folds
can be obtained with the calculation of a  probability density function
(PDF)~\cite{Linsefors:2013cd,Bolliet:2017czc}, which can be performed
with initial conditions defined during the bounce~\cite{Ashtekar:2011rm}
or even in a contraction phase before the bounce~\cite{Chen:2015yua}. In Refs.~\cite{Shahalam:2017wba,Li:2018fco,Sharma:2018vnv,Li:2019ipm,Shahalam:2019mpw} different potentials were investigated in the context of \textsc{LQC}, including power-law potentials~\cite{Shahalam:2017wba}, monodromy potentials with a modulation term~\cite{Sharma:2018vnv}, alpha-attractor potentials~\cite{Shahalam:2019mpw}, and chaotic and Starobinsky potentials in the framework of modified \textsc{LQC} models~\cite{Li:2019ipm}. The duration of inflation was analyzed in all of these models by setting initial conditions at the bounce surface, providing very interesting results.

In this paper we are interested in obtaining the \textsc{PDF} for the  number
of inflationary \textit{e}-folds in \textsc{LQC} by following the perspective adopted in Refs.~\cite{Linsefors:2013cd,Linsefors:2014tna,Bolliet:2017czc,Martineau:2017sti},
which suggests a natural quantity to which a  flat prior can be
assigned, providing the means to define initial conditions in a consistent
way. {}Following this approach, we will define the set of initial
conditions in the remote past of the contraction phase prior to the
bounce, i.e., when the Universe is classic and well understood.  In
Refs.~\cite{Linsefors:2013cd,Linsefors:2014tna,Bolliet:2017czc,Martineau:2017sti}
studies were made of different forms of the inflationary
potential, with the initial conditions taken far back in the contracting
phase including only the energy density of the inflaton as the main
ingredient of the early Universe and at the bounce.

The present paper extends the analysis performed in
Refs.~\cite{Linsefors:2013cd,Bolliet:2017czc,Martineau:2017sti,Shahalam:2017wba,Li:2018fco,Sharma:2018vnv,Li:2019ipm,Shahalam:2019mpw} by considering higher powers of the monomial potential and analyzing the duration of inflation with a Higgs-like potential as a function of the vacuum expectation value (\textrm{VEV}). These analyses provided us with a great comparison tool for the second part of our work, where we consider radiation as an additional ingredient of the energy density budget  around the bounce, which is done for the first time. There are many good reasons for including radiation in these studies. First, it is not excluded at all that prior to inflation the Universe could have been radiation dominated. In fact, radiation has been claimed to be an important ingredient in setting appropriate initial conditions for inflation~\cite{Bastero-Gil:2016mrl}. Dissipative effects are
naturally expected in the early Universe, where radiation can be produced either by decaying processes involving the  inflaton field through its coupling to other fields or through other fields not
directly coupled to the inflaton. These processes---which can also lead to reheating at the end of cold inflation as the inflaton oscillates around its minimum---are similarly expected  to occur in the pre-bounce phase, deep in the contracting phase, where the inflaton also displays oscillations. In fact, initial conditions in the
contracting phase with inflaton oscillations are exactly the initial conditions advocated in
Refs.~\cite{Linsefors:2013cd,Bolliet:2017czc,Martineau:2017sti}.  In addition, radiation production may not even need strong breaking of adiabaticity caused by the inflaton oscillations but can also happen under quasiadiabatic conditions. An outstanding example of this is radiation production processes happening in the warm inflation picture~\cite{Berera:2008ar} (for earlier studies of warm inflation in the context of LQC 
see, for example, Refs.~\cite{Zhang:2013yr,Herrera:2010yg,Herrera:2014mca,Graef:2018ulg,Benetti:2019kgw}).  
There are also many other possible sources of radiation, including gravitational particle production
mechanisms~\cite{Parker:1968mv,Parker:1969au}. In particular,
gravitational particle production has been shown to be very efficient in the bounce phase of several
models~\cite{Quintin:2014oea,Tavakoli:2014mra,Haro:2015zda,Hipolito-Ricaldi:2016kqq,Celani:2016cwm,Scardua:2018omf,Zago:2018huk} and we also expect the same to happen in LQC, as recently shown in Ref.~\cite{Graef:2020qwe}.  The presence of radiation may adversely affect the predictions for inflation in LQC, and this provides the main motivation for the present work.

This paper is organized as follows. In Sec.~\ref{TB} we briefly review the theoretical background about \textsc{LQC} and explain how radiation can be included in the system. In Sec.~\ref{LQCphases} we describe the different dynamic regimes expected in LQC, from the deep contracting phase prior to the bounce, up to the slow-roll phase in the expanding regime. In Sec.~\ref{analysis} we describe the method used in our analysis  and give the results obtained therein. In
Sec.~\ref{additional} we discuss  additional effects  neglected in our analysis that could contribute to the results. {}Finally, in Sec.~\ref{conclusion} we give our conclusions.

\section{Theoretical Background}
\label{TB}

In this section we briefly review the background dynamics of LQC.  We also discuss the generality of the inflationary phase that can be
generated in LQC and how to obtain the most likely number of
inflationary \textit{e}-folds of a given model.

In LQC  cosmological models are described using  LQG principles. As
discussed in Ref.~\cite{Ashtekar:2011rm}, in LQC the spatial geometry
is encoded in a variable $v$ proportional to the physical volume of a
fixed, fiducial, cubic cell, in place of the scale factor $a$, i.e.,
 \begin{equation}\label{volume}
     v = - \frac{4\,{\cal V}_{0}\,a^{3}\,M^{2}_{\rm Pl}}{\gamma},
 \end{equation}
where ${\cal V}_{0}$ is the comoving volume of the fiducial cell,
$\gamma$ is the Barbero-Immirzi parameter obtained from the
calculation of the black hole entropy in LQG (the typically value adopted in LQC is $\gamma \simeq 0.2375$ \cite{Meissner:2004ju}), and
$M_{\rm Pl} \equiv 1/\sqrt{8 \pi G}= 2.4 \times 10^{18}\,GeV$ is the reduced Planck mass. The conjugate momentum to $v$ is denoted by $b$ and it is given by 
 \begin{equation}\label{conjmomentum}
    b = - \frac{\gamma P_{(a)}}{6\,a^{2}\,{\cal V}_{0}\,M^{2}_{\rm
        Pl}}, 
 \end{equation}
where $P_{(a)}$ is the conjugate momentum to the scale
factor. Therefore, the pair $(v,b)$ is used in place of
$(a,P_{(a)})$. These variables are related by the Poisson bracket $\{v, b\}= -2$. After solving the Einstein equations, $b$ is related to the Hubble parameter via $b= \gamma H$.

We are interested in  the Friedmann equation modified in
LQC. Hence, let us consider the equation of motion for $v$, which is given by~\cite{Ashtekar:2011ni}
\begin{equation}\label{motionvolum}
    \dot{v}= \frac{3}{\gamma \lambda} v \sin ({\lambda b}) \cos
    ({\lambda b}),
\end{equation}
with $\lambda$ given by
\begin{equation}\label{constant1}
    \lambda^{2} = \frac{\sqrt{3}\,\gamma}{2 M^{2}_{\rm Pl}}.
\end{equation}
 LQC modifies the dynamics of the Einstein equations and, in terms of effective LQC solutions, the Hubble parameter can be written as
\begin{equation}\label{hubbleeffective}
    H = \frac{1}{2 \gamma \lambda}\sin({2 \lambda b}),
\end{equation}
where $b$ ranges over $(0 , \pi/\lambda)$, and in the limit $\lambda\to 0$ GR is recovered. The energy density $\rho$ is related to the LQC variable $b$ through
\begin{equation}\label{density}
    \frac{\sin^{2}(\lambda b)}{\gamma^{2}\lambda^{2}} = \frac{\rho}{3
      M^{2}_{\rm Pl}}.
\end{equation}
Thus, by combining the Eqs.~\eqref{density} and \eqref{hubbleeffective} the Friedmann equation in LQC assumes the form~\cite{Ashtekar:2011rm}
\begin{equation}\label{friedmann}
    \frac{1}{9}\,\left(\frac{\dot{v}}{v}\right)^{2} \equiv H^{2} =
    \frac{\rho}{3 M^{2}_{\rm Pl}} \left(1 -
    \frac{\rho}{\rho_{\textrm{cr}}}\right),
\end{equation}

where  $\rho_{\textrm{cr}}=2\sqrt{3}M^{4}_{\textrm{Pl}}/\gamma^{3}$.

Through the modified Friedmann equation~\eqref{friedmann} we can explicitly see the underlying quantum geometric
effects~\cite{Ashtekar:2011ni}, with the singularity replaced by a quantum bounce when $\rho = \rho_{\textrm{cr}}$. For $\rho \ll \rho_{\rm cr}$ we recover GR, as expected. The above expression holds independently of the particular characteristics of the inflationary parameters when
initial conditions for the Universe are assumed.

In a cosmological scenario where the Universe is dominated by the
energy density of a scalar field $\phi$ the inflaton--- the equation of motion for $\phi$ is simply
\begin{equation}
 \ddot{\phi} + 3H \dot{\phi} + V_{,\phi} = 0,
\label{eom}
\end{equation}
where $V_{,\phi} \equiv dV(\phi)/d\phi$ is the field derivative of the
inflaton's potential.  In the present work, we also include radiation
as a main ingredient of the energy density. Radiation can be included
by considering decaying processes involving the inflaton field,
where part of its energy density is converted into radiation and parametrized through a dissipation term in Eq.~(\ref{eom}), with dissipation coefficient $\Gamma$,
\begin{equation}
 \ddot{\phi} + 3H \dot{\phi} + \Gamma \dot \phi + V_{,\phi} = 0,
\label{eomdiss}
\end{equation}
and supplemented by the equation for the evolution of the radiation
energy density,\footnote{Note that in the oscillating regime for the inflaton, we can also replace the term $\dot \phi^2$ in Eq.~(\ref{rhoR}) by its average over an oscillation cycle~\cite{Kolb:1990vq}, $\langle \dot \phi^2 \rangle_{\rm cycle} = \rho_\phi$, which for Eq.~(\ref{rhoR}) gives the more
standard form used, e.g., in reheating studies.}
\begin{equation}
\dot \rho_R + 4 H \rho_R = \Gamma \dot \phi^2.
\label{rhoR}
\end{equation}
Note that by multiplying Eq.~(\ref{eomdiss}) by $\dot \phi$, adding it to Eq.~(\ref{rhoR}), and using that $\rho_\phi= \dot \phi^2/2 + V(\phi)$,
$p_\phi = \dot \phi^2/2 - V(\phi)$ and $p_R=\rho_R/3$, we obtain 
\begin{equation}
\dot \rho_{\rm total} + 3 H ( \rho_{\rm total} + p_{\rm total})=0,
\label{eomfluid}
\end{equation}
which is the usual fluid equation for the total energy density,
$\rho_{\rm total} = \rho_\phi + \rho_R$. This shows explicitly that Eqs.\,(\ref{eomdiss}) and (\ref{rhoR}) are conservative with respect to the total energy density, as expected.

Alternatively to the approach adopted in Eqs.~(\ref{eomdiss}) and (\ref{rhoR}), 
we could also assume radiation to be already present in the system, at some early time, independent of explicitly relying on modifying the dynamical equations by the introduction of decay processes, e.g., directly affecting  the inflaton field in Eq.~(\ref{eomdiss}).
Radiation in this case could be due, for example, to the decay of
other fields at some earlier times, or even through gravitational
particle production mechanisms. In this case, at the time we set the
initial conditions for the inflaton, there already can also be some nonvanishing early radiation energy density. In this work we consider both situations and show that our results remain
unaltered and independent of the details of the radiation production mechanisms that might be at play.  In either case, the total energy density is then given by $\rho = \dot{\phi}^{2}/2 + V(\phi) + \rho_R$, implying the modified Friedmann equation
\begin{equation}\label{zizi}
H^{2} =  \frac{\dot{\phi}^{2}/2 + V(\phi) + \rho_R}{3 M^{2}_{\rm Pl}}
\left[1 - \frac{\dot{\phi}^{2}/2 + V(\phi) +
    \rho_{R}}{\rho_{\textrm{cr}}}\right],
\end{equation}
and its time derivative
\begin{equation}
\dot{H} = -\frac{3\dot \phi^2 + 4\rho_R}{6 M_{\rm Pl}^2} \left[1 -
  2\frac{\dot{\phi}^{2}/2 + V(\phi)+\rho_R}{\rho_{\textrm{cr}}}\right].
\label{dotH}
\end{equation}

\section{Phases of LQC}
\label{LQCphases}
 
Let us divide the dynamics of the Universe in LQC into that prior to and after the bounce.

\subsection{Pre-bounce regime}

Let us consider a sufficient time back in the contracting phase where the inflaton is in an oscillatory regime. In this pre-bounce regime, where $H<0$, $\phi$ and $\dot \phi$ are oscillating  with increasing amplitudes or have damped oscillations, depending on whether the decay processes given by $\Gamma$ in Eq.~(\ref{eomdiss}) are present or absent ($\Gamma =0$). Either way, we can characterize this regime by the conditions
\begin{align}
 \rho \ll \rho_{cr},\quad H < 0, \quad   H^2 \ll |V_{,\phi\phi}|,
 \label{oscil}
\end{align}
and when including $\Gamma$, with also the condition $\Gamma < 2
\sqrt{|V_{,\phi\phi}|}$ such that the inflaton is still oscillating,
albeit in an underdamped way.  {}Following the proposal of
Refs.~\cite{Linsefors:2013cd,Bolliet:2017czc}, we define initial
conditions for the Universe in this phase of an oscillating inflaton field in the contracting phase.  In Ref.~\cite{Linsefors:2013cd} it
was suggested as a natural variable to assign initial conditions in
this regime the phase $\delta$ of the field oscillations. Though
this is a natural choice for the simple case of the quadratic inflaton
potential, where both $\phi$ and $\dot \phi$ have simple oscillating
(or, in the presence of $\Gamma$, underdamped) solutions in the regime
of Eq.~(\ref{oscil}), for other types of potentials the expression for
the field and its derivative in the contracting phase may not be that
simple. Therefore, in our numerical analysis (which we describe below) we will assign initial conditions directly to the
scalar field and its derivative by choosing appropriate values for
the initial density ratio defined by $\alpha=\rho/\rho_{\textrm{cr}}$, with
$\alpha$ sufficiently small such that the conditions of
Eq.~(\ref{oscil}) hold. Note that in the case where $\Gamma=0$, but
still including some initial radiation energy density, this will also
entail some upper bound for the initial radiation energy density.

As we approach the bounce, starting from the point given by
Eq.~(\ref{oscil}),  there might be a phase of {\it slow-roll
deflation}.  This phase is the opposite of what happens in 
slow-roll inflation, as it is still in the contraction phase. This
phase is characterized by an almost constant $\dot \phi$ and a
linearly growing $|\phi|$. The conditions for slow-roll deflation are
\begin{align}
 \rho \ll \rho_{cr},\quad  H < 0,\quad H^{2} \gg |V_{,\phi\phi}|,
 \quad  V(\phi) \gg \dot\phi^{2}/2, \;\; \rho_R.
\end{align}
However, the probability that this phase will occur is  small since
almost none of the possible paths that start at low energy in the
contraction phase have an exponential contraction phase in the
pre-bounce. Thus, the fraction of trajectories that have a significant
contraction phase is very small, implying that  the dynamics of these
trajectories (for a high energy density) are strongly dominated by
kinetic energy~\cite{Bolliet:2017czc}. In the presence of radiation
the  probability of this phase gets even slimmer since, as one gets close
to the bounce, the radiation energy density (which grows faster than
the potential energy density in the contracting regime) will tend to
dominate over $V(\phi)$.

{}Finally, just prior to the bounce, there is a phase of {\it
  superdeflation}.  This phase, which  occurs just before the bounce and thus still in the contracting phase when $H < 0$, lasts from the time
when  $\dot{H}=0$ until $H = 0$ (i.e., already in the bounce). In this phase, we then have 
\begin{align}
H^{2} \gg |V_{,\phi\phi}|, \quad   \dot\phi^{2}/2 \gg V(\phi),\;\;
\rho_R.
\label{superdefl}
\end{align}
We typically find that this phase of superdeflation happens very
quickly, typically lasting  less than a Planck
time~\cite{Zhu:2017jew}. The presence of radiation can make it even
shorter, as the radiation will tend to take a large portion of the
energy density prior to the bounce. 

\subsection{Post-bounce regime}

Immediately after the bounce, if the energy density is mostly
dominated by kinetic energy, we have a phase of {\it superinflation}.
This phase, already at the beginning of the expansion, goes from just
after the bounce  (when $H = 0$, i.e., $\rho=\rho_{\textrm{cr}}$) until the
point  where $\dot{H} = 0$. The conditions for superinflation are
again the same as in Eq.~(\ref{superdefl}), however, at the
commencement of the expanding phase.  This is also a very short phase,
just like the superdeflation one, and radiation also tends to make it
shorter.

After the bounce phase, the kinetic energy quickly decreases as $\dot
\phi^2 \propto 1/a^6$ and the radiation decreases as $\rho_R \propto
1/a^4$, while the potential energy density $V(\phi)$ only changes slowly. The inflaton dynamics after the bounce and throughout the
preinflationary phase is just monotonic, with no oscillations~\cite{Zhu:2017jew}; thus, we expect no significant radiation production in this phase.
By also neglecting other possible sources of radiation in this phase, the potential energy of the inflaton will  eventually dominate the
energy content of the Universe and the standard slow-roll inflationary
phase will set in, but with a duration that can be strongly affected
by the radiation present already in the earlier phases, 
as we will see in the next section.  

At the
beginning of slow roll we have that $\rho \ll \rho_{\textrm{cr}}$, the quantum
corrections to the Friedmann equation are negligible, and the
cosmological equations reduce to the usual ones of GR. Let us estimate
the number of \textit{e}-folds of expansion from the bounce to the beginning of
slow-roll inflation. In the absence of radiation, the transition from
the stiff matter kinetic-energy-dominated regime after the bounce to the
slow-roll phase happens rather quickly, with the equation of state
changing from $w\approx 1$ to $w \approx -1$ typically in
less than one \textit{e}-fold~\cite{Zhu:2017jew}. Depending on the amount of the radiation present, we can have an intermediate
radiation-dominated regime~\cite{Graef:2018ulg,Graef:2020qwe}
where the equation of state at the bounce $w\simeq 1$ changes to $w\simeq 1/3$, before assuming the value $w\simeq -1$ at the start of inflation (which occurs when the equation of state becomes smaller than $-1/3$). 

The number of \textit{e}-folds during the preinflationary phase $N_{\textrm{preinfl}}$, from the
bounce to the start of slow roll, can be
approximately estimated in the absence of radiation by assuming that 
around the start of slow roll, at time $t_{\textrm{sr}}$,  $\rho_{\rm
  kin}(t_{\rm sr})\equiv \dot \phi^2(t_{\rm sr})/2 \sim \rho_V(t_{\rm sr})$, where
$\rho_V \equiv V(\phi)$.  By also recalling that the bounce is
dominated by the kinetic energy, $\rho_{\rm kin}(t_{\rm bounce})
\simeq \rho_{\rm cr}$, then, we have that
\begin{equation}
\rho_{\rm kinetic}(t_{\rm sr})\simeq \frac{ \rho_{\rm cr}}{a^6 (t_{\rm
    sr})} \sim \rho_V(t_{\rm sr}).
\label{tsr}
\end{equation}
As an estimate for $\rho_V(t_{\rm sr})$ we can use the upper bound
obtained by the \textit{Planck} data on the scale of inflation when the pivot
scale exits the Hubble radius~\cite{Akrami:2018odb}, $\rm V_*<(1.6\times
10^{16}{\rm GeV})^4$. Using this result in Eq.~(\ref{tsr}), we obtain
\begin{equation}
N_{\rm preinfl}= \ln\left[\frac{a(t_{\rm sr})}{a(t_{\rm
      bounce})}\right] \sim \frac{1}{6}
\ln\left(\frac{\rho_{\rm cr}}{\rm V_*}\right) \sim 4.3.
\label{Npre}
\end{equation}
Note that the estimate given by Eq.~(\ref{Npre}) is based on the value
for the scale of inflation at around the time that the relevant wavelengths
cross the Hubble radius during inflation, which happens at around $60$
or so \textit{e}-folds before the end of inflation. {}For inflation lasting much
longer than the minimum, we do not expect a much higher value for the
potential at the beginning of inflation as a consequence of the
slow-roll conditions.  As we will explicitly see for the different
inflation models studied in the next section, despite the fact that each model predicts rather different values for the total number of \textit{e}-folds of
inflation, we always find that $N_{\rm preinf} \sim 4$. This shows
that the estimate given by Eq.~(\ref{Npre}) is quite satisfactory when
in the absence of radiation. The effect of radiation on the above
estimate can be understood by the fact that it removes part of the
energy density of the inflaton that would otherwise be
available. Thus, it delays the start of inflation and $N_{\rm
  preinfl}$ increases when  compared to the cases when radiation is
absent. This effect will be explicitly seen in our numerical results.
This result can also be understood analogously in terms of the scale of inflation
in Eq.~\eqref{Npre}. Radiation not only delays the start of inflation, but also
decreases $\rm V_*$, thus increasing the estimate for $N_{\rm preinfl}$.

\section{Method, numerical strategy, and results}
\label{analysis}

As already mentioned, in this work we closely follow the
procedure suggested in Refs.~\cite{Linsefors:2013cd,Bolliet:2017czc}
to obtain the appropriate PDFs for the expected number of \textit{e}-folds of
inflationary expansion for the different models that we will
analyze. The procedure can be summarized by the following steps:  
\begin{itemize}
\item  We consider an appropriate initial time deep in the
  contracting regime prior to the bounce. The initial energy density
  $\rho_0$ is such that $\rho_0 = \alpha \rho_{\rm cr}$ is small enough
  ($\alpha \ll 1$) so as to start the evolution early in the
  contracting phase with the inflaton field in the oscillatory regime
  defined in Eq.~(\ref{oscil}).  {}For all of our numerical studies we consider in particular that $\alpha < 8 \times 10^{-17}$,
  while checking  the consistency of the results for each potential as
  $\alpha$ was varied.
\item {}For the considered initial energy density $\rho_0$ at the
  initial time $t_0$, we take random samples of initial values for the
  scalar field, which will be localized around the minimum of its
  potential with some dispersion $\Delta \phi$, such that $
  -\phi_0-\Delta \phi\leq \phi(t_0) \leq \phi_0+\Delta \phi$, where
  $\phi_0$ is the value of the inflaton field at the bottom of its
  potential. The radiation energy density can either be introduced
  through dissipative processes like in Eqs.~(\ref{eomdiss}) and
  (\ref{rhoR}), starting with $\rho_R(t_0)=0$ with a fixed dissipation
  coefficient $\Gamma$, or we can set an initial radiation energy
  density $\rho_R(t_0) \neq 0$ and vanishing  dissipation coefficient,
  as explained in the previous section. {}Finally, the time derivative
  of the inflaton field is then set as $\dot \phi(t_0) = \pm \sqrt{2}\sqrt{\rho_0-V(\phi(t_0))-\rho_R(t_0)}$, with a randomly chosen sign.
\item We solve the dynamics with the produced initial conditions from
  the contracting branch to the end of slow-roll inflation in the
  expanding branch using the dynamical equations of motion given by
  Eqs.~(\ref{eomdiss}), (\ref{rhoR}), and (\ref{dotH}), which are solved
  for the different inflationary models described by the potential
  $V(\phi)$. In the cases studied with radiation being produced in the
  contracting phase due to the inflaton's oscillations, we assume
  perturbative decay analogously to what can happen in the reheating
  phase after inflation~\cite{Abbott:1982hn,Albrecht:1982mp}, setting
  $\Gamma=0$ when the inflaton stops oscillating, which happens right
  after the bounce. Due to the very short duration of the bounce phase 
  ($\Delta t \sim t_{\rm Pl}$), we neglect any source of particle production 
  during the bounce. Therefore, we can set $\Gamma=0$ just after the bounce 
  in the expanding phase. In a second approach, for comparison, we simply 
  consider the presence of an already present initial amount of radiation 
  energy density in the contracting phase at the beginning of our simulations
  and set $\Gamma=0$ in  Eqs.~(\ref{eomdiss}) and (\ref{rhoR}), and then evolve the
  system from the initial time $t_0$ to the end of inflation with the resulting 
  equations.
\item {}For each initial condition sampled we obtain the corresponding
  number of \textit{e}-folds and produce the associated PDF, from which the
  appropriate statistical analysis can be performed.  We work
  with samples ranging from $1000$ to $5000$ points for each model
  analyzed, which we find to be enough to obtain satisfactory
  statistics.
\end{itemize}

\subsection{Models}

In this work we study two classes of inflation models with
primordial potentials as given below.

\subsubsection{Power-law monomial potentials}
\label{chaotic}

In this class of models, we have $V(\phi)$ given by 
\begin{equation}\label{powerlawV}
V= \frac{V_0}{2n} \left( \frac{\phi}{M_{\rm Pl}}\right)^{2n},
\end{equation}
and we explicitly analyze the cases for the quadratic, quartic, and
sextic forms of the potential (corresponding to the powers $n=1,\,2,\;
{\rm and}\; 3$, respectively).  The model given by
Eq.~(\ref{powerlawV}) covers the class of inflationary models
corresponding to large-field models~\cite{Lyth:2009zz}.

\subsubsection{The Higgs-like symmetry-breaking potential}

The Higgs-like symmetry-breaking  potential is given by the following expression:
\begin{equation}\label{higgsV}
V=V_{0} \left[1 - \left(\frac{\phi}{v}\right)^2 \right]^{2},
\end{equation}
where $v$ denotes the  \rm VEV of the field. The Higgs-like symmetry-breaking  potential can represent either a
small-field inflation model if inflation starts (and ends) at the
plateau part of the potential (i.e., for $|\phi| < |v|$), or a large-field model, for which inflation ends in the chaotic part of the
potential ($|\phi| > |v|$). Throughou our analysis with this potential, we explicitly distinguish these two possibilities and produce results
for both.

In all of the above potentials the constant $V_{0}$ is obtained from
the normalization of the CMB spectrum, and this is how we define $V_0$
for each of the above potentials. {}For definiteness, we have fixed
$V_0$ for each model as $V_0/M_{\rm Pl}^4 \simeq 3.41 \times 10^{-11}$
for the quadratic monomial potential, $V_0/M_{\rm Pl}^4 \simeq 1.37
\times 10^{-13}$ for the quartic monomial potential, and $V_0/M_{\rm Pl}^4 \simeq 1.82 \times 10^{-16}$ for the sextic monomial
potential.  Note that for the Higgs-like symmetry-breaking  potential Eq.~(\ref{higgsV}), the normalization of the spectrum implies that
the value of $V_0$ will also have a dependence on the VEV of the
inflaton, but for the \rm VEV's we consider, $14 M_{\rm Pl}
\leq v \leq 25 M_{\rm Pl}$,  $V_0$ has values ranging from
$V_0/M_{\rm Pl}^4 \simeq 1.72 \times 10^{-14}$ to $3.82 \times
10^{-14}$.\footnote{Note that, depending on the decay processes and the amount of radiation at the time that the CMB scales leave the Hubble radius during inflation, the normalization $V_0$ can change with respect to the vacuum values, as in, e.g., the case in warm inflation~\cite{Ramos:2013nsa}. However, we do not consider these processes that can change the primordial power spectrum in the present study when fixing the value of $V_0$.}

Note that  the monomial potentials like the ones we consider
here are already ruled out in the simple scenarios of cold inflation,
according to the  {\it Planck} results~\cite{Akrami:2018odb}.  The
Higgs-like potential, on the other hand, can still be compatible with the observations for
some ranges of the {\rm VEV}.  However, when radiation processes are present (most notably as is the case for these models when studied in the warm inflation context) 
all of these potentials can be shown to agree with the observations (see, e.g.,
Refs.~\cite{Ramos:2013nsa,Bartrum:2013fia,Bastero-Gil:2016qru,Benetti:2016jhf,Benetti:2019kgw,Bastero-Gil:2019gao}).
Looking ahead at the possibility of extending the analysis presented
here to warm inflation, this is why we consider the above
potentials in particular, besides, of course, the fact that they are
well motivated in the context of particle physics models in general.

\subsection{Results}
\label{results}

Having explained the numerical strategy that we employ in
our analysis, we now give the corresponding results obtained by using
each of the primordial inflaton potential models defined by
Eqs.~(\ref{powerlawV}) and (\ref{higgsV}).  {}For comparative
purposes, we first consider the case where radiation is absent
throughout the evolution, from the contracting phase at the initial
time $t_0$ to the end of inflation, and then consider explicitly how
radiation influences these results.

\subsubsection{Results in the absence of radiation}

\begin{center}
\begin{figure}[!htb]
\includegraphics [width=7.5cm]{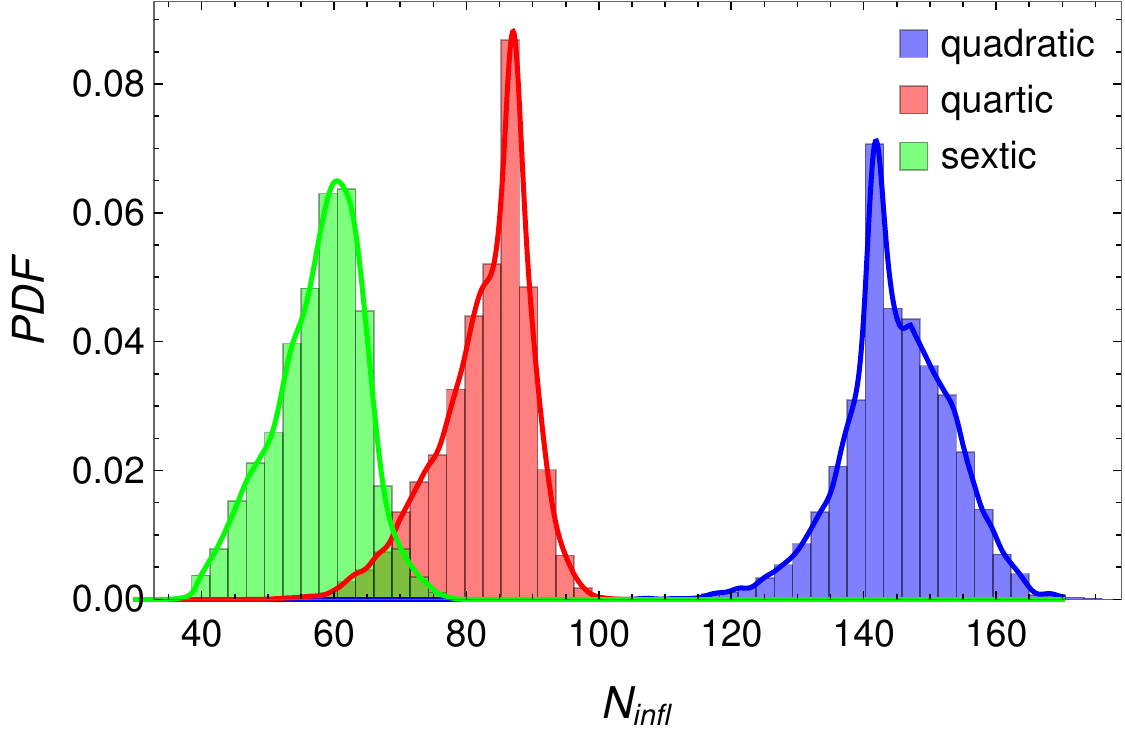}
    \caption{PDF for the total number of inflationary \textit{e}-folds for
      the monomial power-law potentials in LQC  obtained when
      radiation is neglected throughout the evolution.}
    \label{fig1}
\end{figure}
\end{center}

In  {}Fig.~\ref{fig1} we show the PDFs obtained for the total number
of inflationary \textit{e}-folds for the three cases considered for the monomial
power-law potential (\ref{powerlawV}), i.e., for the quadratic
($n=1$), quartic ($n=2$), and sextic ($n=3$) potentials.

As we see from {}Fig.~\ref{fig1}, as we increase the power $n$ of the
potential the number of \textit{e}-folds decreases. The PDFs
for the three cases considered have a dispersion of around $20$ \textit{e}-folds
from the peak of the  distribution, and quickly vanish at the
extrema. In particular, we obtain no more than about a total of
80 \textit{e}-folds of inflationary expansion for the sextic potential.  One
recalls that from the results for the perturbation spectra in LQC, one
typically requires at least around 80 \textit{e}-folds of total expansion from
the bounce in LQC to the end of inflation, such that the quantum
effects on the primordial power spectra are sufficiently
diluted~\cite{Zhu:2017jew}. On the contrary, if the total expansion
lasts less than this minimum, the LQC effects on the spectra would already
be visible. As the
preinflationary expansion that starts from the bounce until the
beginning of inflation does not last more than about 4 \textit{e}-folds (see
discussion at the end of Sec.~\ref{LQCphases} and also the explicit
results on this given below), this already puts the sextic potential in strong tension with
the observations and excludes all other higher-power monomial potentials ($n > 3$) when considering the predicted number of \textit{e}-folds
alone in LQC, even when these models are implemented in the warm
inflation picture.\footnote{Also, in standard cold inflation scenarios
  the monomial power-law potentials are strongly disfavored based on
  the values for the tensor-to-scalar ratio and/or the spectral tilt
  predicted by them~\cite{Akrami:2018odb}.} On the other hand, the
quartic potential (and all other cases with $n < 3$) can most easily
satisfy the required minimum amount of expansion from the bounce to
the end of the inflationary phase. {}Finally, we note that the result
we have obtained for the quadratic potential, which gives a  $N_{\rm infl}$ of around $140$, is in agreement with the
previous results already obtained in Ref.~\cite{Linsefors:2013cd} for
this specific form of the inflationary potential. The results for the
quartic and sextic forms of the potential are new.

\begin{center}
\begin{figure}[!htb]
\includegraphics[width=7.5cm]{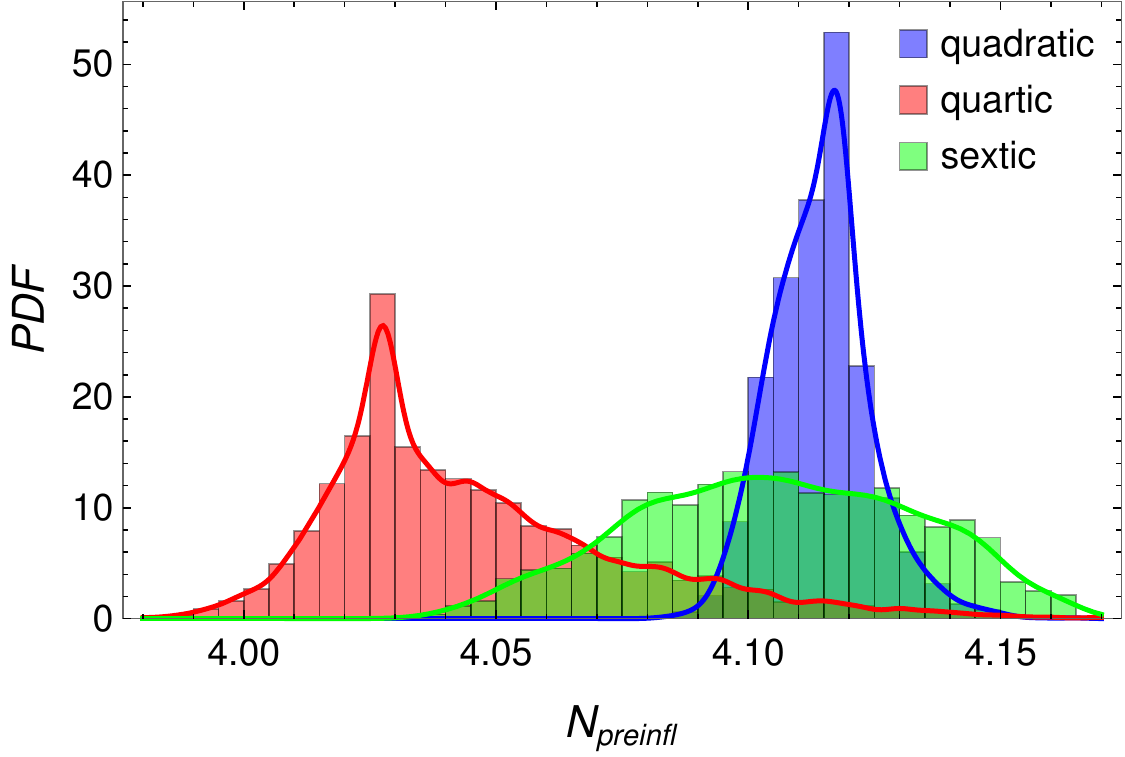}
    \caption{Number of preinflationary \textit{e}-folds for the power-law
      potentials in LQC.}
    \label{fig2}
\end{figure}
\end{center}

To complete our analysis for the monomials power-law potentials, in
{}Fig.~\ref{fig2} we also show the results for the PDFs for the number
of preinflationary \textit{e}-folds, which considers the expansion from the
bounce to the beginning of the slow-roll inflation.  We note from the
results shown in {}Fig.~\ref{fig2} that, despite the differences in the
PDFs, the expected number of preinflationary \textit{e}-folds is $N_{\rm
  preinfl} \sim 4$ for all three models, which agrees with the
estimate given by Eq.~(\ref{Npre}).

\begin{center}
\begin{figure}[!htb]
\subfigure[]{\includegraphics[width=7.5cm]{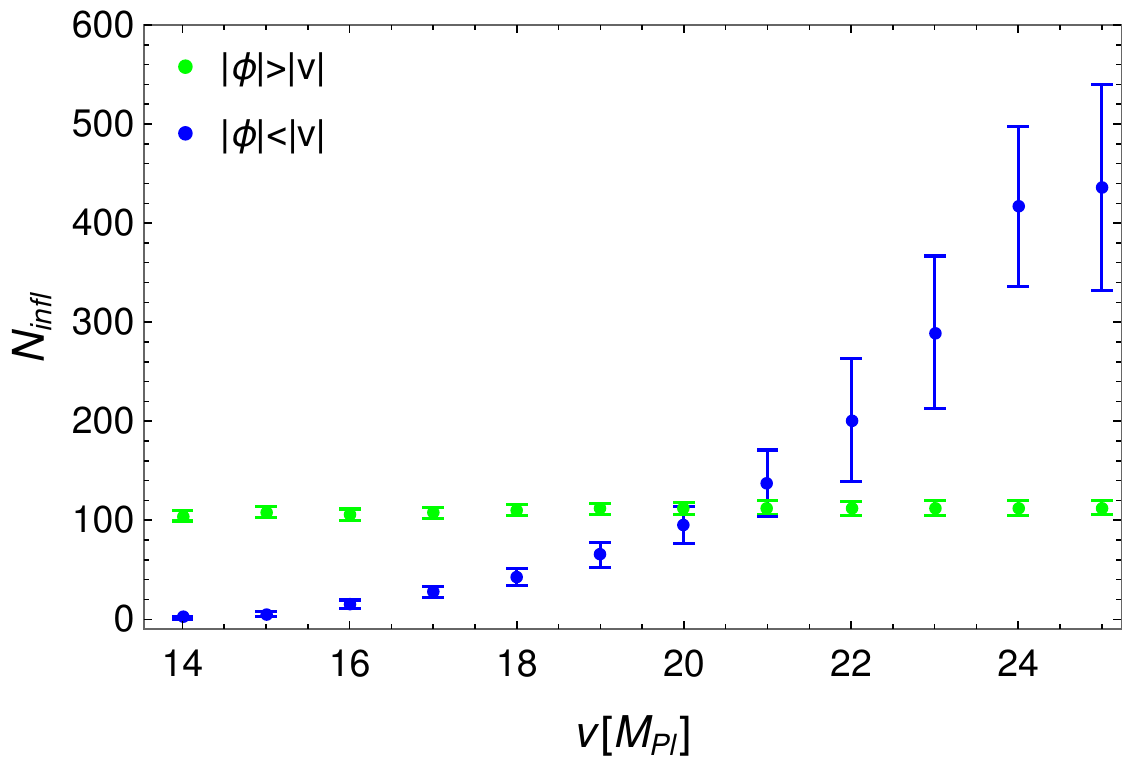}}
\subfigure[]{\includegraphics[width=7.5cm]{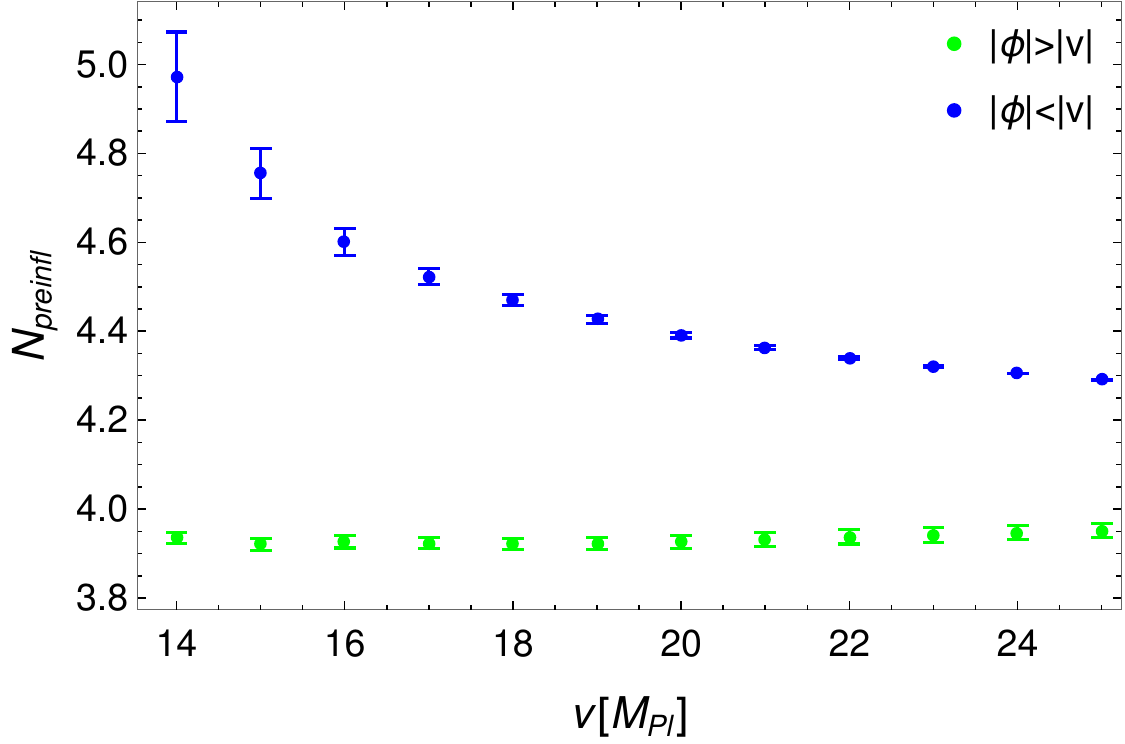}}
\caption{(a)Number of total inflationary \textit{e}-folds and (b) number of preinflationary \textit{e}-folds for the Higgs-like symmetry-breaking  potential in LQC as a function of the {\rm VEV}. The errors bars in the plots indicate the $1\sigma$ standard deviation of the results from the median obtained from the respective PDFs. All cases were analyzed without radiation in the evolution.}
\label{fig3}
\end{figure}
\end{center}

{}For the Higgs-like symmetry-breaking  potential (\ref{higgsV}) we analyze cases for different values of the VEV $v$.  The
results for the total number of \textit{e}-folds of inflation as a function of
$v$ are summarized in {}Fig.~\ref{fig3}(a). Note that we have
explicitly separated the cases of inflation happening in the plateau
part of the potential ($|\phi| < |v|$) from the cases of inflation
happening in the chaotic part ($|\phi| > |v|$).  We observe that the
number of \textit{e}-folds in the chaotic part of the potential is consistently slightly
 above 100 \textit{e}-folds for the cases shown in
{}Fig.~\ref{fig3}(a). But we have also verified that when $|v|\lesssim
8 M_{\rm Pl}$ (not shown in {}Fig.~\ref{fig3}) the expected $N_{\rm
  infl}$ starts to approach the one seen for the quartic potential in
the monomial case, as expected.  We have also analyzed whether there
would be any preference for inflation happening in either part of the
potential.  However, the results of our simulations do not show a
significant preference for inflation to occur in the plateau or 
chaotic part of the potential. The probability for a given initial
condition to end up leading to inflation in the plateau or chaotic
region of the potential is always around 50$\%$, with a slight oscillation
around this value as $v$ is changed.  But the results do show that for
$|v|\lesssim 14 M_{\rm Pl}$ there are essentially no more initial
conditions leading to inflation starting and ending in the plateau
region. {}Furthermore, for $|v|\lesssim 19 M_{\rm Pl}$ the expected
number of \textit{e}-folds in the plateau part of the potential is already
smaller than around 80 \textit{e}-folds, and the discussion given above regarding the monomial potentials with $n\gtrsim 3$ applies here as well.  

We note that inflation in the plateau region is subject to the
well-known initial condition problem (see, e.g.,
Ref.~\cite{Bastero-Gil:2016mrl} and references therein). In
particular, the smaller the VEV in the Higgs-like potential, the
less of an attractor the slow-roll trajectory becomes. Interestingly
enough, in our results this initial condition problem for inflation in
the plateau does not manifest in the number of initial conditions
ending up in the plateau region, but instead in a reduction of the
total number of inflationary \textit{e}-folds as $v$ decreases. On the other
hand, the larger the VEV, the larger the number of \textit{e}-folds in the
plateau region, which here is a manifestation of the increase of the
attractor nature for the slow-roll trajectories on the plateau and as the plateau
gets flatter as $v$ increases, hence leading to potentially more \textit{e}-folds.   
In {}Fig.~\ref{fig3}(b) we give the results for the predicted number of
preinflationary \textit{e}-folds for the Higgs-like potential. Once
again, we have explicitly separated the cases of initial conditions
leading to inflation in the plateau or in the chaotic parts of the
potential. The results show that $N_{\rm preinfl}$ decreases with $v$
for the case of inflation occurring in the plateau and tends to
converge towards $N_{\rm preinfl} \sim 4.3$ for $|v| > 24 M_{\rm Pl}$.
On the other hand, for inflation occurring in the chaotic part of the
potential, we obtain that $N_{\rm preinfl}$ is almost independent of
$v$, though the data shows a slow increase as $|v|$ increases and
$N_{\rm preinfl}$ is slightly below 4, but still consistent with the
estimate given by Eq.~(\ref{Npre}).

\begin{center}
\begin{figure}[!htb]
\subfigure[]{\includegraphics[width=7.5cm]{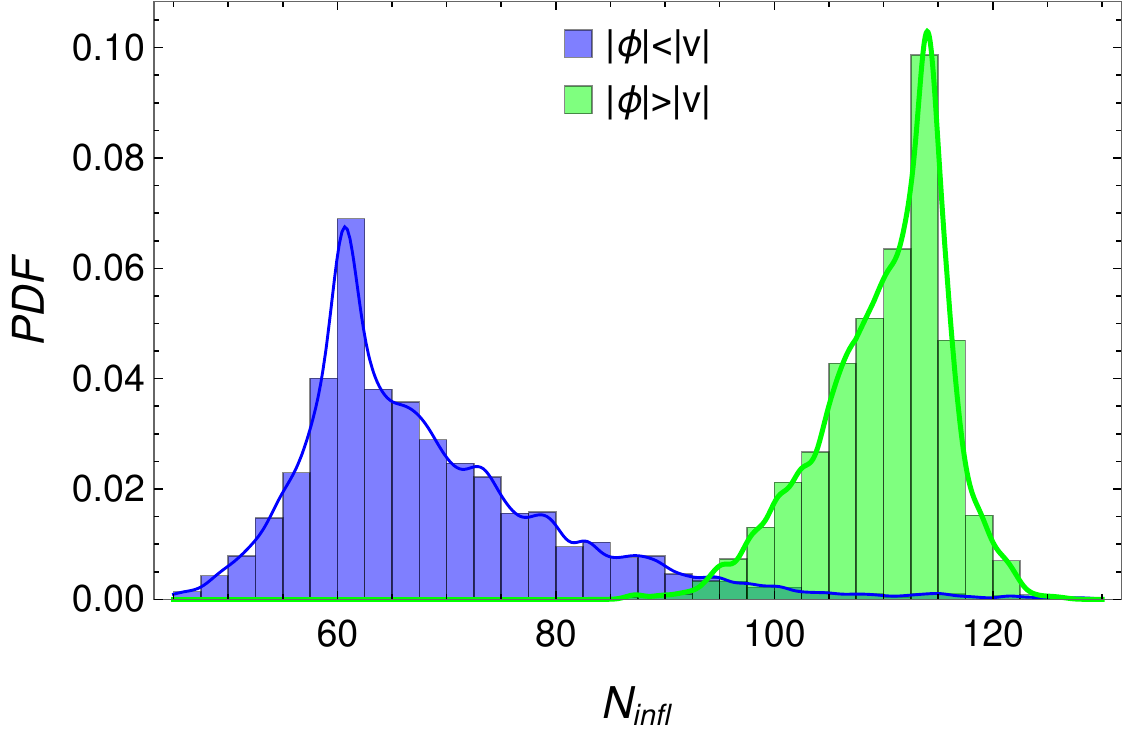}}
\subfigure[]{\includegraphics[width=7.5cm]{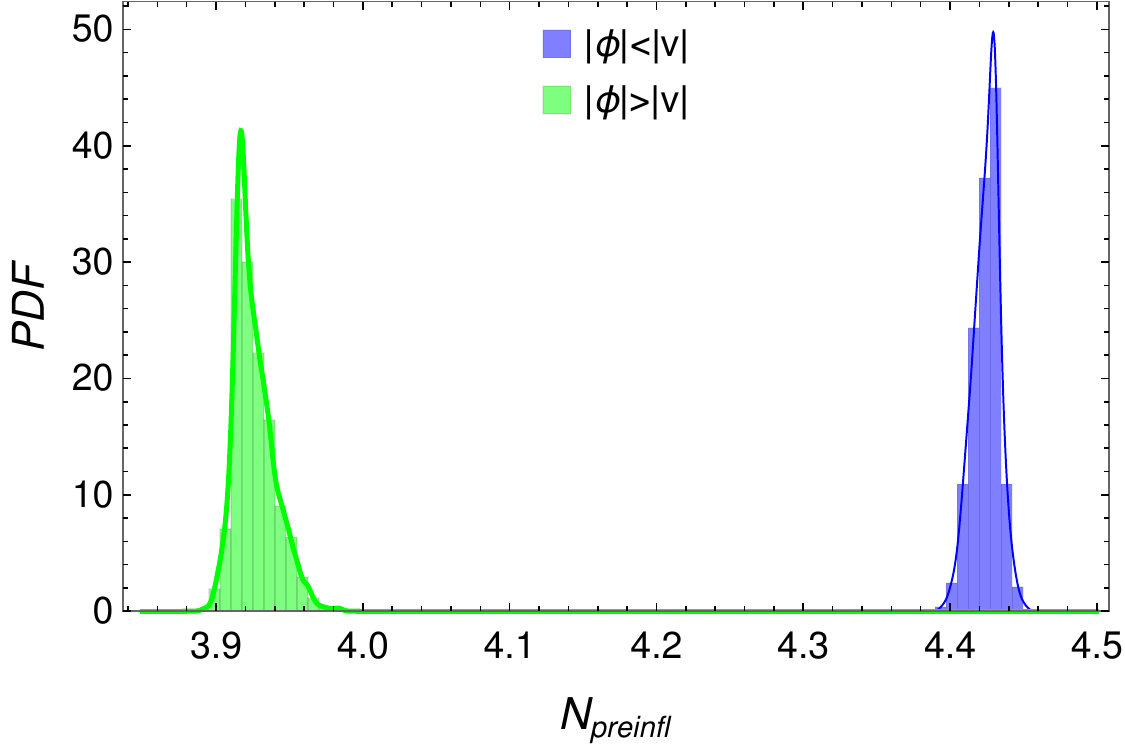}}
\caption{(a)PDF for the number of inflationary \textit{e}-folds for the chaotic and plateau parts of the Higgs-like symmetry-breaking potential in LQC considering the value $v=19 M_{\rm Pl}$. (b)PDF for the number of preinflationary \textit{e}-folds for the chaotic and
  plateau parts of the Higgs-like potential in LQC considering the
  same VEV. As in the previous figures, radiation is absent throughout
  the evolution.}
\label{fig4}
\end{figure}
\end{center}

As a complement and example case extracted from the above results for
the Higgs-like symmetry-breaking  potential, in {}Fig.~\ref{fig4}(a)
we explicitly show the PDF for the number of inflationary \textit{e}-folds, taking as an example the vacuum expectation value  of the Higgs-like
symmetry-breaking  potential to be $v=19 M_{\rm Pl}$. Likewise, in 
{}Fig.~\ref{fig4}(b) we also show the PDF for the number of
preinflationary \textit{e}-folds from the bounce to the beginning of the
slow-roll inflation obtained for the same VEV.

\begin{center}
\begin{table}[!htb]
\caption{Values for the median and standard deviation ($1\sigma$) for the number of 
preinflationary and inflationary \textit{e}-folds for the power-law and Higgs-like symmetry-breaking  potentials in LQC  in the absence of radiation effects.}
\label{tab1}
\begin{tabular}{c|cc}
\hline\hline
 & \multicolumn{2}{c}{Median and Standard Deviation}\\
Model & $N_{\rm preinf}$ &  $N_{\rm infl}$ \\
\hline
Quadratic & $4.115 \pm 0.010$ & $144 \pm 8$\\
Quartic & $4.038 \pm 0.030$ & $84 \pm 7$ \\
Sextic & $4.10  \pm 0.06$ & $59  \pm 7$ \\
Higgs ($v=19 M_{\rm Pl}$) & & \\
plateau  & $4.426 \pm 0.009$ & $65 \pm  13$\\
Higgs ($v=19 M_{\rm Pl}$) &  & \\
chaotic & $3.923 \pm 0.014$ & $111 \pm 6$ \\
\hline\hline
\end{tabular}
\end{table}
\end{center}

{}Finally, for completeness we summarize our main results that can be
extracted from all of the PDFs in the Table~\ref{tab1}, where we give the
results for the median and standard deviation for $N_{\rm infl}$ and
$N_{\rm preinfl}$ for each of the models studied when neglecting
radiation effects. {}For the Higgs-like symmetry-breaking  potential,
we only give results obtained from the specific example shown in
{}Fig.~\ref{fig4}. {}For the other VEVs studied, see
{}Fig.~\ref{fig3}.

\subsubsection{Results in the presence of radiation}

Let us now study how the inclusion of radiation affects the above
results. We start by considering Eqs.~(\ref{eomdiss}), (\ref{rhoR}), and (\ref{dotH}) with the dissipation coefficient
$\Gamma$. One notes that here $\Gamma$ parametrizes a radiation
production process where part of the energy density of the inflaton is
converted to radiation. As already pointed out in the previous
section, there can be many other different processes at play
generating radiation that are not directly related to the inflaton (e.g., the decay of spectator fields, gravitational particle production,
etc.). Parametrizing radiation production like the perturbative decay of
the inflaton might represent only one such process. However, as explained
below, our results are only dependent on the amount of radiation prior
to the bounce and much less on which particular process
(or processes) might lead to it. This significantly simplifies our study, in addition to
showing that our results should not be sensitive to the details of
the dynamics of radiation production in the contracting phase. These are rather strong claims, and we justify them by considering as an example the case of the monomial quadratic inflaton potential.

\begin{center}
\begin{figure}[!htb]
\subfigure[]{\includegraphics[width=7.5cm]{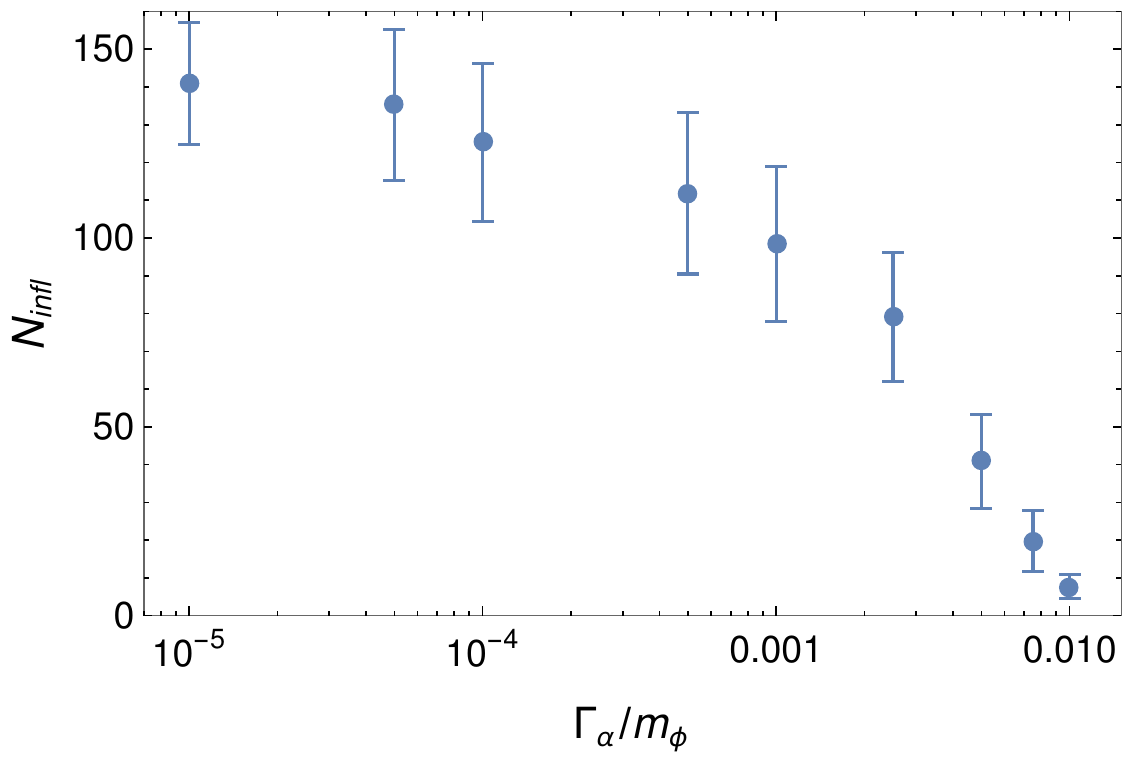}}
\subfigure[]{\includegraphics[width=7.5cm]{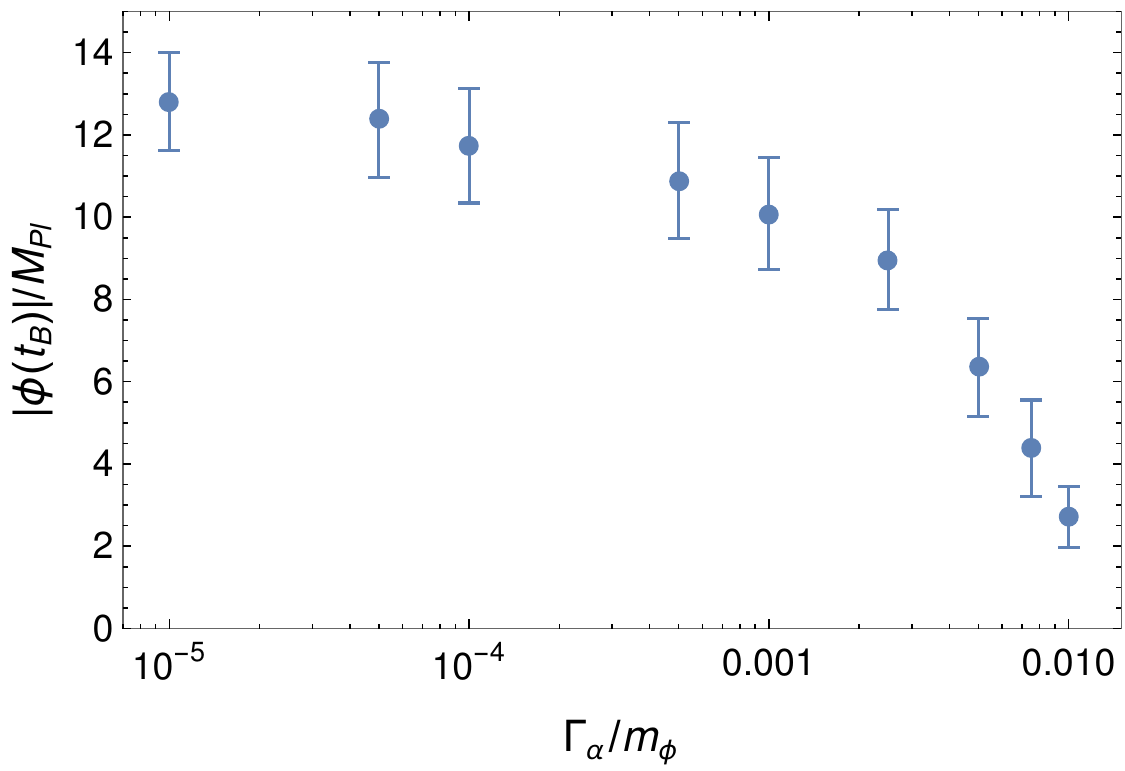}}
\subfigure[]{\includegraphics[width=7.5cm]{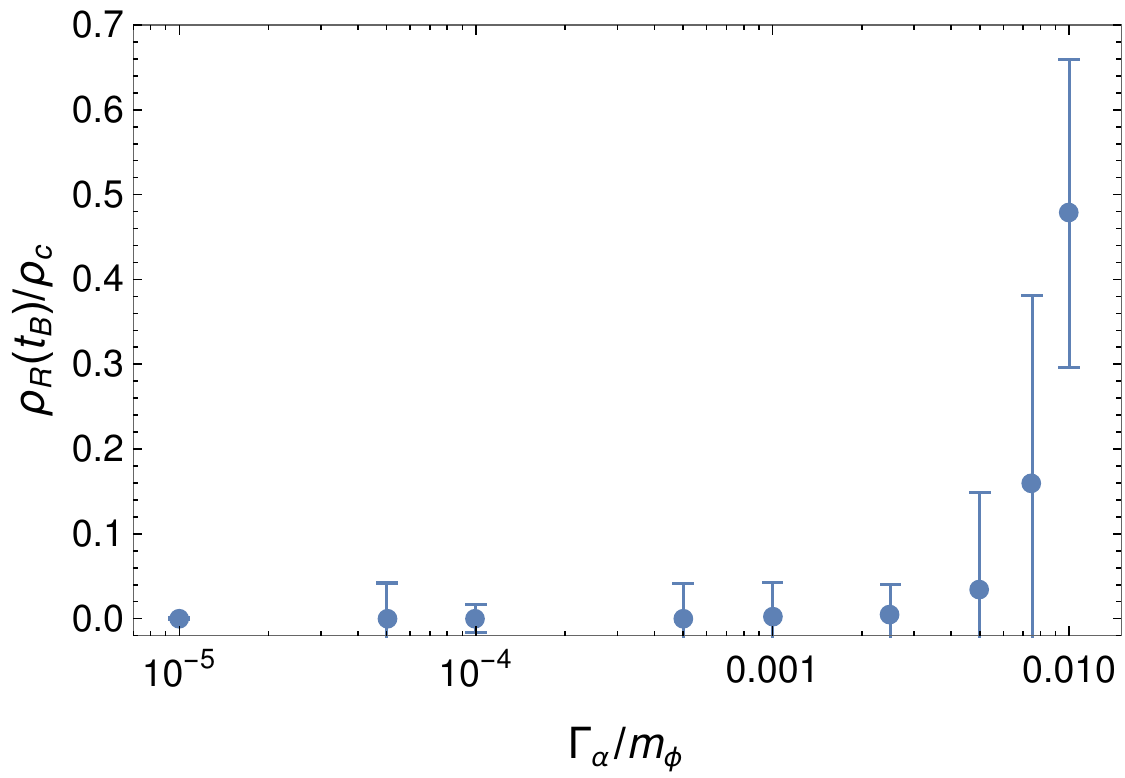}}
\caption{(a)Number of total inflationary \textit{e}-folds, (b)  modulus of the amplitude of the inflaton at the bounce, and (c) radiation energy density fraction at the bounce as a
  function of the dissipation rate $\Gamma$, for the case of the
  monomial quadratic inflaton potential. The inflaton mass here is
  given by $m_\phi = V_0^{1/2}/M_{\rm Pl}$. The errors bars in the
  plots indicate the $1\sigma$ standard deviation of the results from
  the median obtained from the respective PDFs.}
\label{fig5}
\end{figure}
\end{center}

In {}Fig.~\ref{fig5}(a) we show the effect of the radiation
production through $\Gamma$ on the expected number of \textit{e}-folds of
inflation for the monomial quadratic model. The larger the $\Gamma$,
the smaller the number of \textit{e}-folds expected for inflation later in
the expanding region post-bounce. This result can also be correlated with
the expected value for the inflaton field at the bounce time $t_B$,
$\phi(t_B)$, as shown in {}Fig.~\ref{fig5}(b). As seen in
{}Fig.~\ref{fig5}(b), the larger the $\Gamma$, the smaller  the
amplitude of the inflaton field at the bounce, and the smaller the
resulting number of \textit{e}-folds.  Note that the smaller resulting potential
energy density of the inflaton at the bounce cannot be compensated by
a larger kinetic energy, since now part of the total energy density at
the bounce comprising the critical density $\rho_c$ will be in the form of
radiation energy density at the bounce $\rho_R(t_B)$, as can be seen
in {}Fig.~\ref{fig5}(c).

\begin{center}
\begin{figure}[!htb]
\subfigure[]{\includegraphics[width=7.5cm]{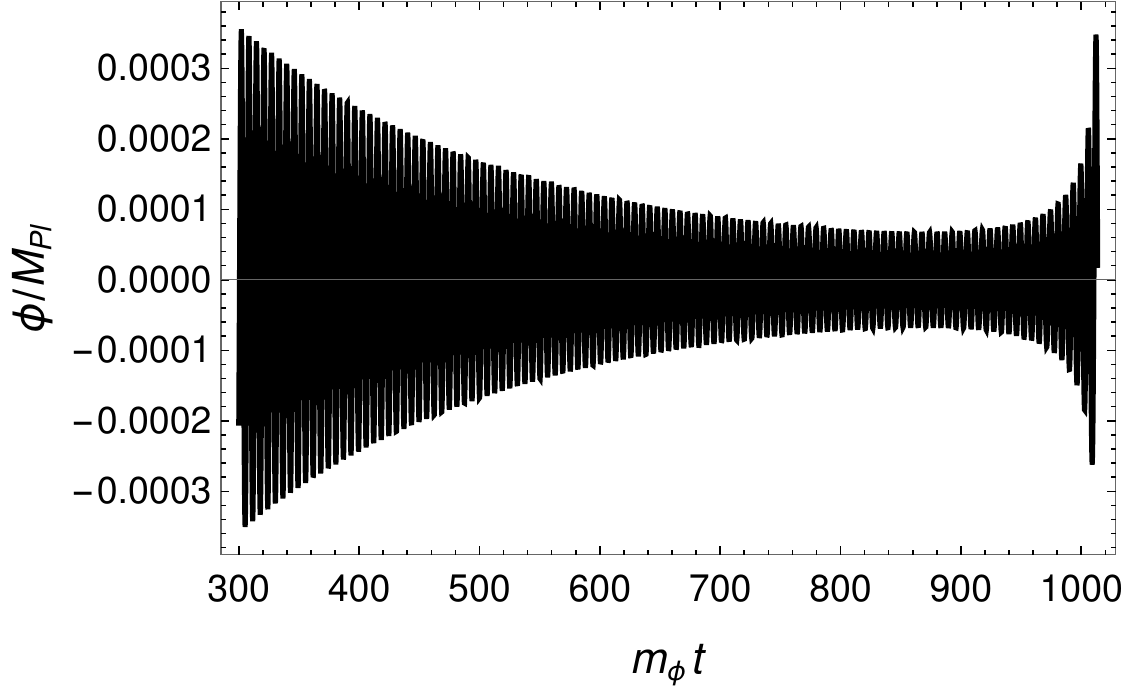}}
\subfigure[]{\includegraphics[width=7.5cm]{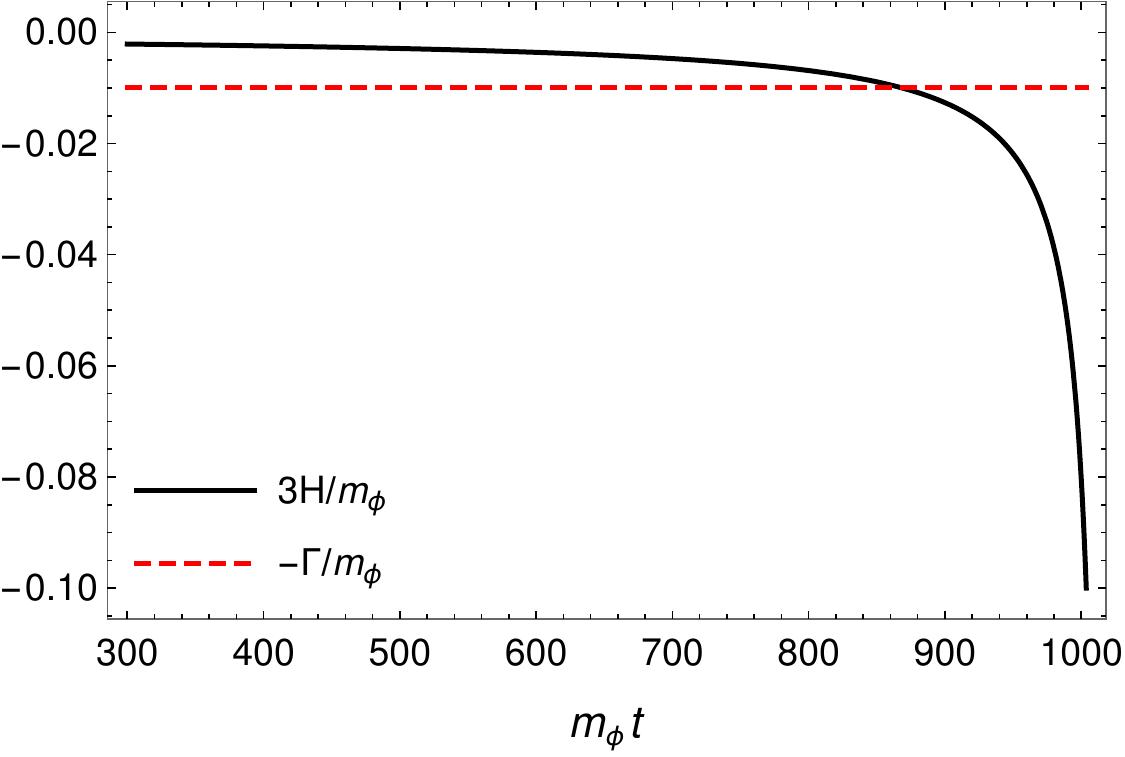}}
\subfigure[]{\includegraphics[width=7.5cm]{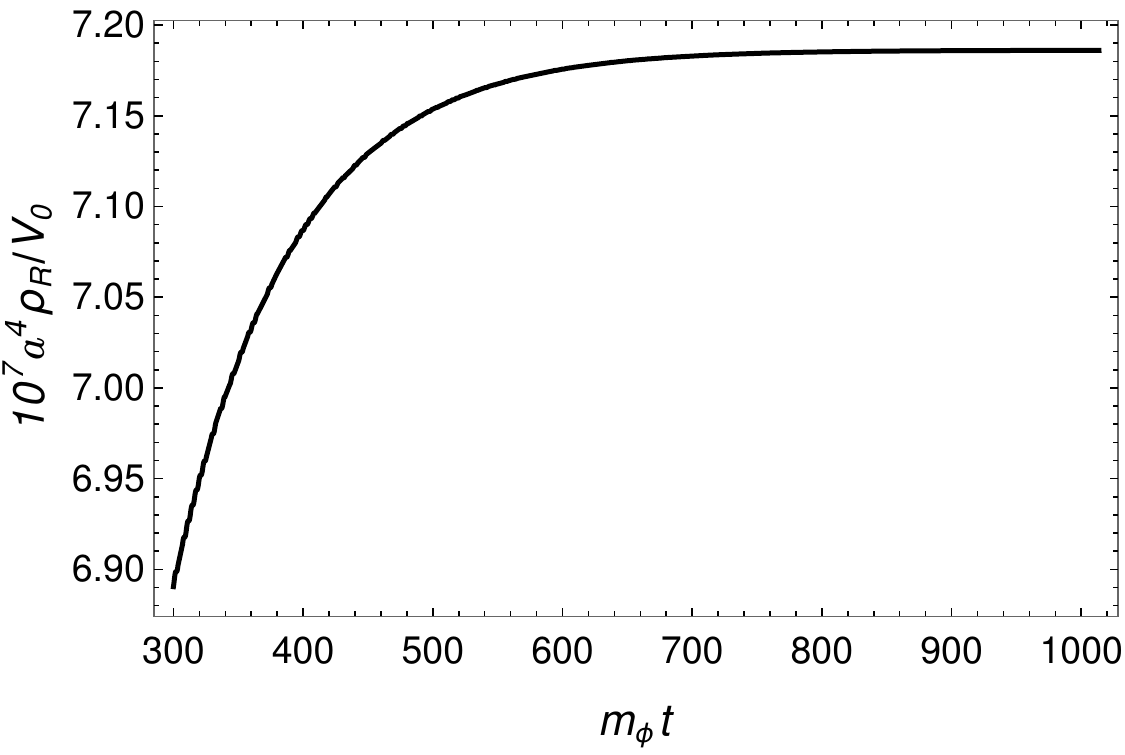}}
\caption{One example of evolutions in the contracting phase, up to
close to (but still below) the bounce instant $t_B$, for (a) the inflaton field, (b) the Hubble parameter, and (c) the radiation energy density times the fourth power of the scale
  factor at the bounce, $a^4 \rho_{\rm R}$, for the case of the monomial
  quadratic inflaton potential. These results were obtained for a dissipation
  rate $\Gamma/m_\phi =0.01$ and a total energy density ratio at
  the initial time given by $\alpha \equiv \rho(t_0)/\rho_c =
  10^{-19}$. Here, the bounce instant is $t_{B} \simeq 1018/m_\phi$.}
\label{fig6}
\end{figure}
\end{center}

As explained in the previous section, these results are obtained from
the PDFs that were generated for different values of $\Gamma$. In
{}Fig.~\ref{fig5} we show the median and $1\sigma$ standard
deviation (shown as error bars) derived from these PDFs. In
this specific example, we consider in particular the fraction
of total energy density at the initial time $t_0$ in the contracting
phase as $\alpha \equiv \rho(t_0)/\rho_c = 10^{-19}$. We have added a
subindex $\alpha$ to $\Gamma$ to explicitly point out that these
results, when expressed in terms of the decay coefficient, should be
interpreted as $\alpha$ dependent.  This is understandable, since
$\alpha$ specifies how far back in the contracting phase we initiate our simulations, and hence determines how many oscillations the inflaton will undergo during its evolution. Of course, the radiation energy density produced will be dependent on this evolution. Thus, for other values of
$\alpha$ we will have a similar behavior as that shown in
{}Fig.~\ref{fig5}, though at different values of $\Gamma$. The
important point to notice is that the Hubble parameter during the
contracting phase increases in modulus (becoming more and more
negative)  before the bounce is approached. Therefore, even if we start the evolution with a $\Gamma > |H|$, at some point before
the bounce we will necessarily have $\Gamma < |H|$. At this point the
inflaton dynamics stops being damped with decreasing oscillations due
to the presence of the dissipation term in Eq.~(\ref{eomdiss}) and
starts to have oscillations with increasing amplitudes. In other
words, the effect of $\Gamma$ on the dynamics is no longer
relevant. In particular, note that radiation production is only
efficient when $\Gamma > |H|$, similarly to what happens in
perturbative reheating, and when $\Gamma  < |H|$
radiation production becomes essentially ineffective. The radiation produced until
that time will then evolve with the metric like $\rho_R \propto 1/a^4$
and increase towards the bounce time, while the inflaton still
oscillates strongly.\footnote{Recall that $|H| < m_\phi$ is the
  condition for the inflaton oscillations, while perturbative decay of
  the inflaton also requires $\Gamma \ll
  m_\phi$~\cite{Kofman:1997yn}.} Note that as we approach the bounce the modification 
of the {}Friedmann equation in  LQC becomes important, and at some point we will again satisfy
the condition $|H| < m_\phi$. However, the time interval of the bounce phase
(when the correction to the {}Friedmann equation is important) is very short, 
typically of the order of a Planck time, such that the production of radiation due to $\Gamma$ is 
negligible during this short period. {}For this reason, we do not need to consider dissipation 
during the bounce phase. In {}Fig.~\ref{fig6}(a) we
explicitly show these expectations for the evolution of the inflaton
field. The evolution of the Hubble parameter in the contracting phase
is shown in {}Fig.~\ref{fig6}(b). Note that when $\Gamma$ drops below
$3|H|$ which in the figure corresponds to the region where the red dashed line ($-\Gamma/m_\phi$) 
is above the black solid line is exactly the time when the damped oscillations of the
inflaton turn into oscillations with increasing amplitudes, just as
expected from Eq.~(\ref{eomdiss}) for the dynamics of the inflaton
field in the contracting phase when $\Gamma=0$.  The resulting radiation energy
density evolution times $a^4(t)$ is shown in
{}Fig.~\ref{fig6}(c). Once again, we see that at the same time that
$\Gamma$ drops below $|H|$, i.e., the inflaton decouples from the radiation,  
the radiation production essentially stops
and $\rho_R a^4 \sim cte$, i.e., the radiation evolves as expected had
we started the evolution at that instant of decoupling $t_{\rm dec}$, with $\Gamma=0$ and with the
given radiation energy density at that instant $\rho_{\rm R}(t_{\rm dec})$ taken as its initial
value. This is why both approaches---i.e., starting evolving the system of
equation in the contracting phase with an explicit dissipation term in
the equations at $t=t_0$ and with $\rho_{\rm R}(t_0)=0$, or simply assuming the
evolution starting at $t_{\rm dec} > t_0$
with an initial nonvanishing radiation
energy density, $\rho_{{\rm R},i} \equiv \rho_{\rm R}(t_{\rm dec})$ at $t_{\rm dec}$,  
but with $\Gamma=0$ turn out to be completely equivalent.

In our systematic analysis of how radiation affects the predictions
for inflation in the models analyzed we still produce the PDFs
starting with initial conditions in the contracting phase with either
radiation being produced through a dissipation term in the evolution
equations, or just assuming an initial radiation energy density  
but setting $\Gamma=0$, as explained above. We have
explicitly checked that the results post-bounce are independent of the
approach used.  In fact, we found that the results are better
presented in a transparent way when they are expressed in terms of the 
fraction of the radiation energy density
that will be present at the time of the bounce, $\rho_{\rm R}(t_B)/\rho_c$.
This way, the results are also expressed in a more general form,
independent of the way the radiation production mechanisms are specified
in the contracting phase. 

\begin{center}
\begin{figure}[!htb]
\subfigure[]{\includegraphics[width=7.5cm]{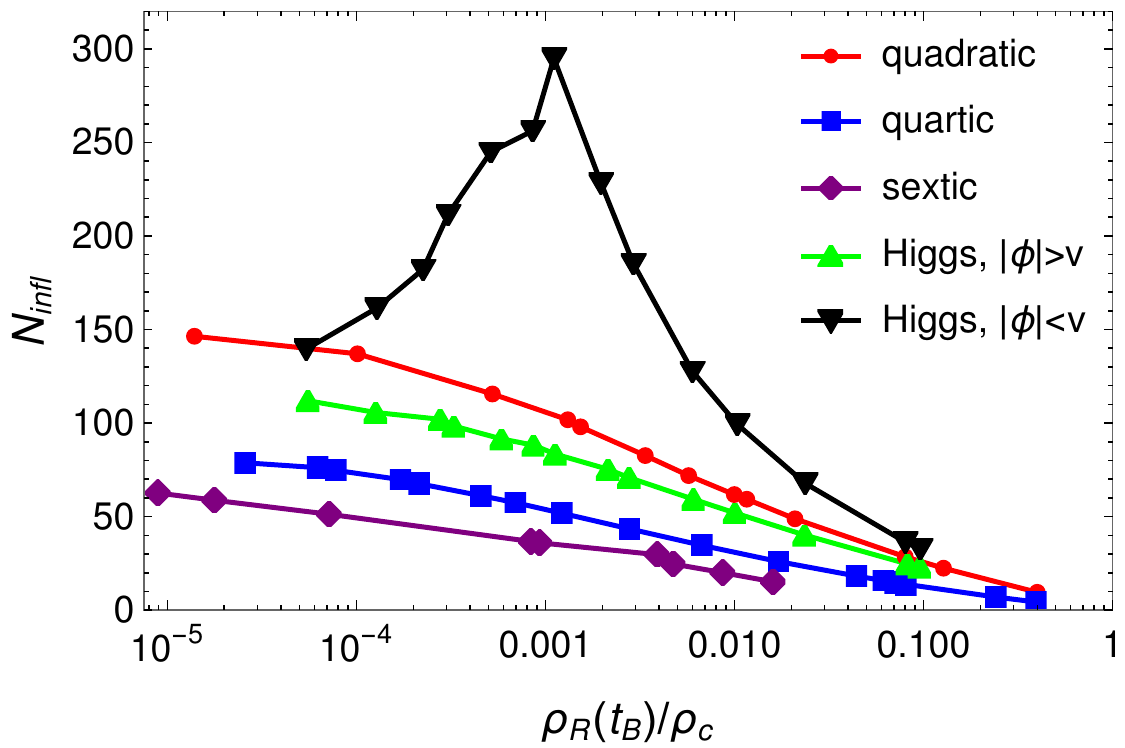}}
\subfigure[]{\includegraphics[width=7.5cm]{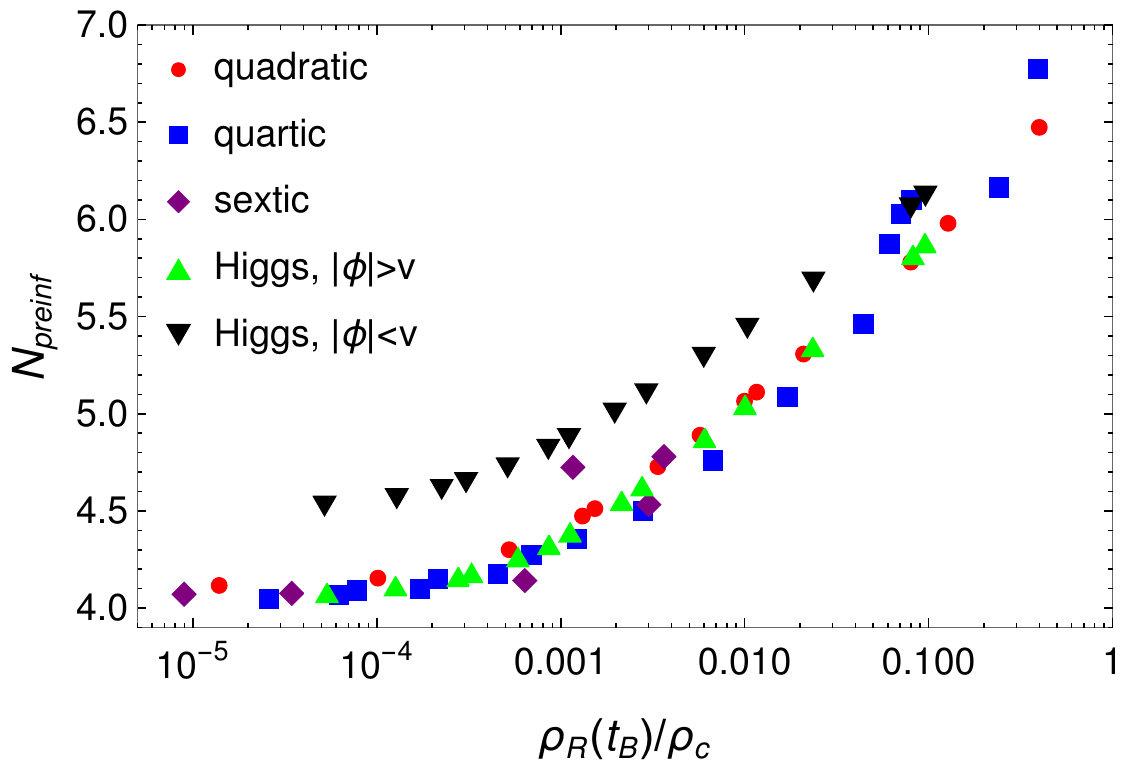}}
\subfigure[]{\includegraphics[width=7.5cm]{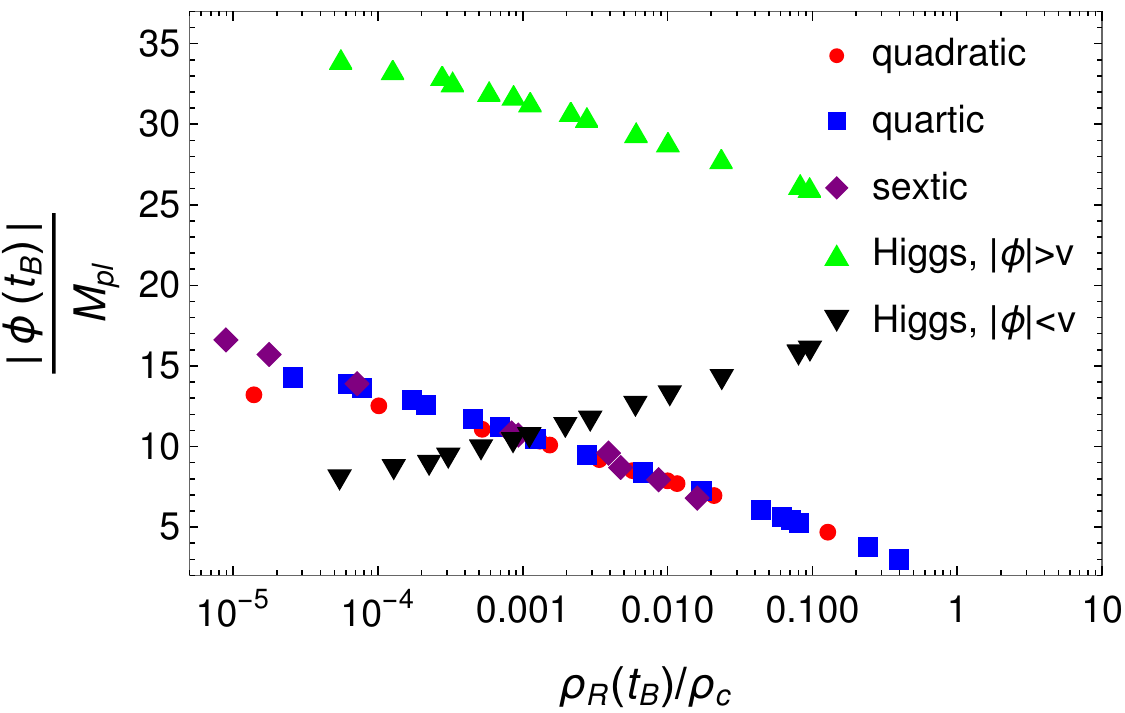}}
\caption{(a)Duration of inflation, (b)duration of the preinflationary phase, and (c)the amplitude for the inflaton at the bounce as a function of the fraction
of the radiation energy density at the bounce. All results refer to the medians extracted 
from the respective PDFs for each of the models studied. The results shown for the Higgs-like
symmetry-breaking potential refer to the case with a VEV $v=21M_{\rm Pl}$.
}
\label{fig7}
\end{figure}
\end{center}

Returning to the results for each of the inflaton potentials
considered in this work and following the procedure explained above,
in {}Figs.~\ref{fig7}(a)-- \ref{fig7}(c) we show the results for the predicted number of
\textit{e}-folds of inflation, the number of preinflationary 
\textit{e}-folds, and the value for the inflaton field amplitude at the
bounce, respectively. To avoid crowding the figures, we do not show the $1\sigma$ standard deviation error bars for each of the data points (obtained from the medians of the respective PDFs for each model).  

Analyzing the results shown in {}Fig.~\ref{fig7}(a), a number of
important features emerge as a result of including the effects of
radiation.  {}For the monomial potential we see the expected effect of
radiation suppressing inflation according to the fraction of radiation
that end up present at the bounce instant $t_B$ and that comes from the earlier evolution
in the contracting phase. In particular, the larger the power $n$
in the monomial potential, the smaller the required fraction of
radiation for the number of \textit{e}-folds of inflation to drop to unsuitable
values to account for the observations. {}For example, for the
quadratic potential the number of \textit{e}-folds drops below 50 when the
fraction of radiation at the bounce is around $2\%$,  for the quartic potential
this fraction is around $0.13\%$, and for the sextic potential it is as
small as $0.0073\%$. In the case of the symmetry-breaking Higgs-like
potential, we have once again explicitly identified the regimes of
inflation happening in the plateau region of the potential, with the
inflaton amplitude at the beginning and end of inflation satisfying
$|\phi| < v$, and the regime of inflation happening in the chaotic
part of the potential, $|\phi| > v$. {}For the example shown in
{}Fig.~\ref{fig7}, we have chosen the case with a VEV $v=21 M_{\rm
  Pl}$, which in the absence of radiation produces approximately the
same number of \textit{e}-folds in the plateau and chaotic parts of the
potential (see, e.g., {}Fig.~\ref{fig3}), which gives $N_{\rm infl} =
118\pm 21$ and $118\pm 6$ for the expected number of \textit{e}-folds for the
plateau and chaotic parts of the potential, respectively. Thus, this particular VEV is better suited for comparative purposes to see the effects of radiation 
on the inflation dynamics when happening in one of the two branches of the potential. 
The behavior of $N_{\rm infl}$ as a function of radiation for the chaotic part of
the potential exhibits a similar trend as the monomial
potentials. It monotonically decreases with the amount of radiation that
permeates the bounce and becomes less than 50 \textit{e}-folds when the fraction
of radiation at the bounce is around $1\%$. However, the behavior for
the number of \textit{e}-folds when inflation happens in the plateau region is
quite peculiar.  It instead shows a growing behavior with the increase of radiation up to
a maximum value, and then decreases. This peculiar behavior can be
explained by the fact that radiation takes up not only potential
energy of the inflaton that it would otherwise have at the bounce
instant, but also kinetic energy. There is then an increased chance for
the initial conditions at the start of the slow-roll inflation to land
close to the top of the potential, thus increasing the number of
\textit{e}-folds. However, as the radiation increases further beyond some
value, the decrease in kinetic energy of the inflaton leads to less
and less initial conditions reaching the top of the potential
plateau, thus decreasing the number of \textit{e}-folds. However,
compared to the other cases we do see that inflation on the plateau is more
resilient to an increase in radiation. The number of \textit{e}-folds of
inflation, for this particular VEV, only drops below $50$ when
the fraction of radiation at the bounce is larger than around $5\%$.

In {}Fig.~\ref{fig7}(b) we see that the number of \textit{e}-folds for the
preinflationary phase increases with the fraction of radiation energy
density. This behavior was already observed before in
Refs.~\cite{Graef:2018ulg,Graef:2020qwe} in the case of the quartic
potential. Here we confirm that this is also a generic expectation for
other forms of primordial inflaton potentials and it can be explained
through the estimate for $N_{\rm preinfl}$ given in the previous
section Eq.~(\ref{Npre}). The presence of radiation will tend to
lower the scale of inflation and, consequently, increase $N_{\rm
  preinfl}$.  {}Furthermore, we see from the results in
{}Fig.~\ref{fig7}(b) that there is a certain universality of the
results for the different potentials.  The data points for the
monomial potentials, along also the Higgs-like potential with
inflation in the chaotic part of the potential, they all group
together, thus having very similar behavior on how $N_{\rm preinfl}$
depends on the radiation energy density fraction at the bounce
instant. In the case of the Higgs-like potential for the inflaton and
with inflation happening along the plateau of the potential the
behavior is similar, though shifted with respect to the other
cases. This is also expected [and also should hold for other
VEVs, as seen, for example, in {}Fig.~\ref{fig3}(b)], given the
different energy scales for inflation happening on the plateau or
chaotic side of the potential. 

{}Finally, a similar universality as that seen
in {}Fig.~\ref{fig7}(b) is also seen in {}Fig.~\ref{fig7}(c), where
we show how the (modulus of the) inflaton field amplitude at the
bounce instant $t_B$ varies with the fraction of the radiation energy
density.  Note that all monomial potentials have data grouping
together. The case of the Higgs-like inflation in the chaotic part of the potential is shifted from the monomial potentials by exactly the value of the
VEV. Had we shifted the potential zero to the VEV point, $\phi \to \phi-v$, 
it would also be grouped with the results for the monomial potentials.  Note that
$|\phi(t_B)|$ decreases as the amount of radiation increases, thus
leading to a smaller number of \textit{e}-folds of inflation, consistent with what we see
in {}Fig.~\ref{fig7}(a). {}$|\phi(t_B)|$ on the plateau part of
the potential, it can only increase towards the VEV, thus also
decreasing the number of \textit{e}-folds.

As a final remark concerning the results obtained for the Higgs-like
potential, similarly to the case  studied in the vacuum, we have found
that the presence of radiation does not favor inflation happening
either in the plateau (small-field) or chaotic (large-field) regions
of the potential.  We have essentially a $50$/$50$ chance for some
initial condition taken deep in the contracting phase to land in either
part of the potential during the inflationary slow-roll phase. This is
quite surprising in view of the fact that for inflation along the large-field part of the potential, like with any chaotic type of inflation,
the slow-roll trajectory  is a local attractor in the field
phase space of initial
conditions~\cite{Albrecht:1986pi,Brandenberger:1990wu}. On the other
hand, plateau inflaton potentials are known to suffer from the initial
condition problem and have to be severely fine-tuned~\cite{Goldwirth:1991rj}. Though large {\rm VEVs} for a Higgs-like
symmetry-breaking potential can strongly alleviate this issue of the
initial conditions, we have explicitly verified that the same trend also holds at small {\rm VEVs}, though we are also led to a
smaller number of \textit{e}-folds, as seen in 
{}Fig.~\ref{fig3}(a).  It appears that this issue with small-field
potentials in LQC turns out to manifest in the most likely (and
sufficient) amount of inflation to happen than in a probability of a
certain initial condition to land on either side of the
potential. Surprisingly, as discussed in the case of the results shown
in {}Fig.~\ref{fig7}(a), there are also regimes where radiation ends up
favoring a larger number of \textit{e}-folds along the plateau part of the
potential. (This is somewhat along the lines of the study done in
Ref.~\cite{Bastero-Gil:2016mrl} showing how a preinflationary phase
dominated by radiation might end up favoring inflation by helping to
localize the inflaton close to the plateau region of the potential.).

\section{Additional effects and future directions}
\label{additional}

It is important to discuss some issues that were not considered
explicitly in this work but could lead to interesting effects.
{}First, in order to make the analysis as general as possible, we
did not consider any specific mechanism for the radiation production.

As discussed in the previous sections, we had simply assumed some {\it a
priori} a particle decay process that leads to radiation production and acts
in the contracting phase.
That the dissipation term is added in the classical regime in the contracting phase
is in particular quite convenient from a quantum field theory perspective in deriving
these dissipation terms. In the classical
regime, quantum gravity effects are negligible and a standard quantum field theory derivation
for dissipation coefficients would apply. The quantum gravity effects would be important
closer to the bounce. However, as explained in the previous section,
the dissipation coefficient $\Gamma$ will in general become smaller than the
(modulus of the) Hubble rate before the bounce is approached in the contracting phase, and from that point
on the radiation production becomes inefficient. Thus, we do not have to deal with
the details of how the quantum gravity effects would affect the radiation production
(at least as far as a quantum field theory derivation for the inflaton dissipation 
coefficient to light fields is concerned). 
We could then think of decay rate terms involving, for instance, explicit interactions
of the inflaton with some light fields, which can be either bosons or
fermions, with interaction Lagrangian densities terms like, e.g., 
${\cal L}_{\rm int}=-g \sigma \phi \chi^2$, with the inflaton coupled
to some other scalar field $\chi$, or ${\cal L}_{\rm int}=-h \phi \bar
\psi \psi $, for the case of couplings to fermions.  Then,  $\Gamma$
refers simply to the  decay processes~\cite{Peskin:1995ev}
(for $m_\phi > 2 m_\chi, 2 m_\psi$) $\Gamma_{\phi\to \chi \chi}= g^2
\sigma^2/(8\pi m_\phi)$ and $\Gamma_{\phi \to \bar\psi \psi}=h^2
m_\phi/(8 \pi)$, respectively, where $g$ and $h$ are two constants. Coupling other fields directly to the
inflaton imposes constraints on the values for the respective
couplings such that quantum corrections coming from these other fields do 
not  spoil the required flatness of the inflaton potential. This
typically requires small coupling constants, $g,\, h \ll 1$, thus
leading to very small decay rates. This in turn would require a long
evolution in the contracting phase such that sufficient radiation can
be produced. However, there are other ways of having light fields (radiation)
coupled to the inflaton and at the same time allowing for large
couplings, provided the inflaton sector is protected by symmetries,
like a shift symmetry in the case where the inflaton is a
pseudo-Nambu-Goldstone boson, as in axionic inflation, or in the
recent constructions involving the inflaton coupled directly to
radiation fields, like in
Refs.~\cite{Bastero-Gil:2016qru,Bastero-Gil:2019gao} in the context of
warm inflation. These processes could also lead to strong dissipation
mechanisms in the contracting phase and possibly be applicable in
the context of the present paper.  Additionally, we could also think
in terms of gravitational particle production. However, these are in general very
inefficient processes  during the oscillatory regime of the inflaton 
in the pre-bounce phase. In this work, we have
also not considered particle production from parametric resonance,
similarly to what might happen in preheating after
inflation~\cite{Kofman:1997yn}, triggered by the oscillations of the
inflaton. Parametric resonance is a very efficient particle production
mechanism  that can cause the energy density of the inflaton to quickly decrease. It would be interesting to investigate how parametric
resonance could manifest itself due to the strong oscillations of the
inflaton in the pre-bounce contraction phase. As we approach the bounce
and the energy density approaches the Planck scale, we might also expect
the opposite behavior to what we would see in the expansion regime
post-inflation, probably with particle fusion happening efficiently,
counterbalancing the evaporation of the inflaton condensate due to its
decay during parametric resonance.  In the high-energy regime close to
the bounce, the energy transfer could then also target the inflaton
field.  Though quite interesting, a full study of the effects would
certainly require a quantum kinetic study of bouncing cosmology in
LQC, something beyond the scope of the present paper.   

We have also neglected in our analysis the possible contribution of inhomogeneities
encoded in the gradient terms, which could be important during the
contraction.  Even though one should not expect these terms to
significantly change the \textsc{PDF}s we obtained, it could be
important to study how these terms could affect the dynamics of the
bounce phase in these models.  In addition, although we have only studied the case of
isotropic LQC, the presence of anisotropies could lead to important
effects. In this context, the analysis made by the authors of
Ref.~\cite{Martineau:2017sti} has shown that considering anisotropic
effects the PDFs can be strongly affected, though we can still draw
predictions from them, like for the number of \textit{e}-folds of inflation. (In
fact, the effects of anisotropies as studied in
Ref.~\cite{Martineau:2017sti} have some similarities to the effects we
have seen here due to radiation. By decreasing the energy density of
the inflaton, we also expect a smaller number of \textit{e}-folds for larger anisotropies.)

Our results can also affect the predictions for each model with respect to
the changes radiation can impose on the power spectrum.
The presence of radiation means that the initial state for which the primordial
scalar curvature perturbations are evaluated is not the Bunch-Davis vacuum,
but likely an excited state for the inflaton. In addition, if the radiation bath
thermalizes, which in general requires that sufficient scattering happens among the
radiation particles, then the formed thermal bath will be carried over into
the preinflationary phase as well. Note that in general we require the condition
that $\Gamma$ be larger than the expansion (contraction) rate of the Universe as a
condition for thermalization~\cite{Kolb:1990vq}. As seen in the example discussed in the previous  
section and shown in {}Fig.~\ref{fig6}(b), this condition is very likely to be
satisfied during some time in the contracting phase. Even though the formed thermal
bath can drop out of equilibrium after $\Gamma$  goes below $|H|$ before the bounce,
the temperature of the thermal bath will simply evolve with the scale factor
as $T \propto 1/a$ from that time onwards and be carried over into the post-bounce
phase, even if no further particle/entropy production happens later on 
and before inflation. The presence of a thermal bath will lead to an enhancement
of the power spectrum~\cite{Bhattacharya:2005wn} and, consequently, to an 
enhancement of the power at the largest scales, i.e., for the smallest 
wave numbers. At the same time, the modification of mode functions due to the presence
of radiation leads to a lowering of the quadrupole moment~\cite{Wang:2007ws,Das:2014ffa,Das:2015ywa}.
In LQC, the primordial scalar curvature power spectrum has also been shown 
to be modified~\cite{Agullo:2013ai,Zhu:2017jew}, also causing an enhancement of the
power at low multipoles. A recent study of these issues in the context of warm
inflation~\cite{Benetti:2019kgw} has shown how these different effects might
counterbalance, easing the lower bound on the duration of inflation determined,
e.g., in Ref.~\cite{Zhu:2017jew}. The results we have obtained in the present paper
certainly call for a more detailed computation of the power spectrum
in LQC whenever radiation might be present in the preinflationary phase.

\section{Conclusions}
\label{conclusion}

Based on the proposal introduced by the authors of
Ref.~\cite{Linsefors:2013cd} on how some well-defined predictions can
be made concerning the probability and duration of inflation in LQC,
we have extended their analysis for other power-law monomial
potentials, like the quadratic, quartic, and sextic potentials, and for the
Higgs-like potential for the inflaton. In the latter model, we
also investigated the results obtained for different values of the
vacuum expectation value.  While in the context of cold inflation the
three power-law potentials are disadvantaged by the {\it Planck}
data~\cite{Akrami:2018odb}, warm inflation can rehabilitate them again
due to the radiation production effects and this justifies using these
potentials in the present study. Besides, as simple potential models,
it is important to consider them for comparison purposes in general.
Motivated by the warm inflation picture, where radiation can be
present throughout the inflationary regime, in this work we 
investigated the effects of radiation on the predictions for inflation
in LQC for all of the above-mentioned primordial inflation potential
models.  

{}Following the procedure detailed in
Refs.~\cite{Linsefors:2013cd,Bolliet:2017czc}, we obtained
different PDFs for different relevant quantities including, for example, the
number of \textit{e}-folds of inflation, the number of preinflationary \textit{e}-folds from
the LQC bounce to the start of the slow-roll inflation, and the fraction
of radiation energy density at the bounce, and drew statistical
conclusions from them for each of the models studied here.  We assumed
initial conditions for the energy density in the remote past, well
before the bounce, and evolved them considering also the radiation.
{}For the cases studied and for the  analysis performed for each of the resulting
PDFs, we found that the number of \textit{e}-folds of the preinflationary phase is
approximately {\rm 4} \textit{e}-folds in all of the models analyzed, and increases
with the radiation energy density. On the other hand, the number of inflationary
\textit{e}-folds changes a lot between the models and also strongly depends on the
radiation energy density present at the bounce time. 

As already explained in previous studies (see, e.g., Refs.~\cite{Martineau:2017sti,Bolliet:2017czc}),
the approach of taking the initial conditions in the classical regime in the contracting
phase leads to very different results than the other approach usually considered in the literature,
i.e., taking the initial conditions at the bounce time.
The reason for this difference can be understood as follows. In general, taking the initial conditions at the bounce time leads to a much larger number of {\it e}-folds, and a prediction for the duration of inflation is
harder to obtain. This is understandable, since if we consider initial conditions at the
bounce, i.e., where $\rho_{\rm total} = \rho_{\rm cr}$, we are allowed in principle to 
consider any value for the inflaton field amplitude up to the value for which 
$V(\phi) = \rho_{\rm cr}$, thus potentially leading to a very large number of {\it e}-folds.
However, by taking initial conditions in the classical regime in the contracting phase,
the amplitude of the inflaton at the bounce is always constrained and the bounce 
is essentially kinetic energy dominated, thus leading to a much smaller number of {\it e}-folds and allowing us to make predictions about the duration of the inflation.
As explained, e.g., in Ref.~\cite{Martineau:2017sti}, this is because a long deflation regime
in the contracting regime (and before the bounce is reached) is strongly suppressed. (In fact,
in all of our numerical simulations and for the different models we have considered, none reached such a regime
of a long deflation.) This then prevents the inflaton from reaching large amplitudes and, consequently, the number
of {\it e}-folds cannot be too large and remains constrained. We have seen this explicitly
in all of our results. 

We obtained that, among the power-law potentials analyzed, the sextic
model in LQC is the one that predicts the lowest value for the number
of inflationary \textit{e}-folds $N_{\rm infl}$, implying a  small
probability of being consistent with the CMB data. The quartic potential,
on the other hand, predicts the most likely $N_{\rm infl}$ to be
around $80$, in the absence of radiation, which suggests a very good
possibility of leading to observable signatures from LQC in the CMB spectrum ~\cite{Zhu:2017jew}. {}For the quadratic model, the
most likely $N_{\rm infl}$ is around $140$, in the absence of
radiation, in agreement with the results obtained in
Ref.~\cite{Linsefors:2013cd}. With such high values of $N_{\rm infl}$,
the effects from the quantum regime would probably be diluted to an
unobservable level whenever there is no radiation present to
affect the dynamics of expansion and  the inflaton.  {}For the
Higgs-like symmetry-breaking  potential we have shown that $N_{\rm
  infl}$ grows with the vacuum expectation value ($v$) for the case of
inflation occurring in the plateau (small-field) region, while for
inflation occurring in the chaotic (large-field) part of the potential
$N_{\rm infl}$ is almost independent of $v$, being always around
$N_{\rm infl} \sim 100$ in the absence of radiation effects. However, radiation has a strong influence on the number of \textit{e}-folds in the plateau
region of the potential. Instead of tending to suppress the duration
of inflation in the plateau, it initially favors an increase of
$N_{\rm infl}$, which can be by a large factor depending on the VEV
and the available radiation energy density. This effect has been
identified as a result of the fact that radiation production decreases the energy
that would otherwise be available for the inflaton (both potential and
kinetic energy). By having a smaller kinetic energy, the inflaton can
then be better localized along the plateau and, hence, increase the
duration of inflation.  

We have also discussed the possible effects that the presence of a radiation 
bath might have on the primordial scalar curvature power spectrum in LQC, 
which also motivates further study in that direction.

\section{Acknowledgments}

 L.N.B. acknowledges financial support of the Coordena\c{c}\~ao de
 Aperfei\c{c}oamento de Pessoal de N\'{\i}vel Superior (CAPES) - Finance
 Code 001. L.L.G. acknowledges financial support of the Conselho
 Nacional de Desenvolvimento Cient\'{\i}fico e Tecnol\'ogico (CNPq),
 Grant No. 307052/2019-2. R.O.R. is partially supported by research
 grants from CNPq, Grant No. 302545/2017-4, and Funda\c{c}\~ao
 Carlos Chagas Filho de Amparo \`a Pesquisa do Estado do Rio de
 Janeiro (FAPERJ), Grant No. E-26/202.892/2017.  


\end{document}